\documentclass[preprint,showpacs,preprintnumbers,amsmath,amssymb]{revtex4}

\usepackage{graphicx}
\usepackage{dcolumn}
\usepackage{bm}

\def\eqcm{\: ,}           
\def\eqpt{\: .}
\def\adj{{\phantom{.}}}   
\def\sT{{\scriptscriptstyle T}}
\newcommand{\beqa}{\begin{eqnarray}}
\newcommand{\eeqa}{\end{eqnarray}}
\newcommand{\bls}{\baselineskip}

\newcommand{\be} {\begin{equation}}
\newcommand{\ee} {\end{equation}}
\newcommand{\g}{\gamma}
\newcommand{\sig}{\sigma}

\newcommand{\eps}{\epsilon}

\newcommand{\nn}{\nonumber}

\newcommand{\Pbarslash}{\kern 0.2 em {\bar P}\kern -0.64em /}
\newcommand{\kbarslash}{\kern 0.2 em {\bar k}\kern -0.5em /}
\newcommand{\Deltaslash}{\kern 0.2 em {\Delta}\kern -0.64em /}
\newcommand{\vpslash}{\kern 0.2 em{v^{\prime}}\kern -0.7em /}

\newcommand{\me}[3]{\langle#1\vert\,#2\,\vert#3\rangle}

\newcommand{\td}{\tilde}

\begin{document}
\title{The off-forward Quark-Quark Correlation Function}
\author{Sabrina Casanova}
\affiliation{Los Alamos National Laboratory, Los Alamos, 87545, NM, US}
\altaffiliation{Max Planck fuer Radioastronomie, Bonn, 53121, Germany}
 \email{casanova@lanl.gov}
\date{\today}

\preprint{}

\begin{abstract}
The properties of the non-forward quark-quark correlation function are examined. 
We derive constraints on the correlation function from the 
transformation properties of the fundamental fields of QCD occurring 
in its definition. We further develop a method to construct an ansatz 
for this correlator. We present the complete leading order set of 
generalized parton distributions in terms of the amplitudes of the ansatz. 
Finally we conclude that the number of independent generalized parton 
helicity changing distributions is four.
\end{abstract}


\maketitle
\section{Introduction}
In quantum field theory the non-perturbative nature of composite particles 
is described by matrix elements of operators of the fundamental 
fields of the theory evaluated between the initial and the final state of the 
particle under consideration. Accordingly, in QCD, the information on the 
internal partonic structure of hadrons is entirely given by matrix elements of all 
possible quark and gluon operators evaluated between hadronic 
states, forward and non-forward.

Forward matrix elements of quark and gluon operators, i.e.~operators 
between hadronic states of equal momenta, are investigated in hard 
inclusive processes, while non-forward matrix elements, i.e.~operators 
between hadronic states of different momenta, have been investigated in recent 
years in the context of non-forward high-energy exclusive processes 
such as Compton scattering in the deeply virtual kinematical limit (~DVCS~) 
and hard diffractive vector-meson production. Experimental results~\cite{Adloff:2001cn,Saull:1999kt,
Airapetian:2001yk,Stepanyan:2001sm} 
indicate the feasibility of measuring these processes much more precisely with dedicated experiments in the future.

In this article we will focus on hadronic matrix elements of bilinear quark operators, 
since these are the ones involved in the description of dominant contributions to non-forward hard processes. 
We consider their Fourier transforms, i.e. the quark-quark correlation functions, which contain 
all information concerning the non-perturbative nature of hadrons. From non-forward quark-quark correlation functions, 
for convenience, the so called generalized parton distributions (GPDs) are defined.~\cite{Radyushkin:1997ki,Ji:1998pc}

Following and generalizing the method developed for the ordinary forward parton distributions~\cite{Ralston:1979ys,Mulders:1995dh}, 
we will formulate the most general ansatz for the quark-quark correlation functions starting from general principles, 
and we will then analyze leading order GPDs by projecting the ansatz with different Dirac matrices. In this 
way we will be able to express the GPDs in terms of the amplitudes entering the ansatz and establish a formal method for 
the determination and classification of the independent quark GPDs.

The number of independent GPDs occurring in a given Dirac projection is not self-evident. For instance it has been debated in the 
literature~\cite{Diehl:2001pm,Hoodbhoy:1998vm} whether there are two or four independent quark helicity changing 
GPDs corresponding to the forward transversity distribution. With the method developed in this article 
we will show unambiguously that there are indeed four independent helicity changing GPDs at leading twist.

The outline of the work is as follows: in Section \ref{sec:correlator} we define the non-forward 
quark-quark correlation function and we introduce a new method to build an ansatz for it. We also derive constraints on the 
non-forward quark-quark correlator imposed by the hermiticity properties of 
the quark fields and their well-known behavior under parity and time reversal operations. In Section 
\ref{sec:GPDs} we relate the non-forward correlation function to leading twist quark GPDs by 
tracing the ansatz with various Dirac matrices and by integrating over quark momentum components. Different Dirac structures 
probe different spin properties of the hadrons. Finally, in Section \ref{sec:conclusions} we draw conclusions and 
discuss the outlook for this subject.

\section{The non-forward quark-quark correlation function\label{sec:correlator}}

\begin{figure}[ht]
\begin{center}
\includegraphics[width=6cm] {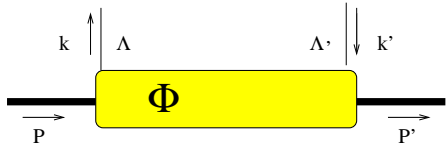}
\end{center}
\caption{\label{fig:correlator}
\sl {Diagrammatic representation of the correlation function 
$\Phi_{i \Lambda', \;j \Lambda }(k,k',P,P^{\,\prime})$. From an initial 
hadron with momentum $P$ a quark with momentum $k$ is taken out, 
and reinserted with momentum $k'=k+\Delta$ to form the final 
state hadron with changed momentum $P^{\,\prime}=P+\Delta$.}} 
\end{figure}

We define the non-forward quark-quark correlation function $\Phi_{i \Lambda', \;j \Lambda}(k,k',P,P^{\,\prime})$, 
depending on the hadron and quark momenta (see Fig.~\ref{fig:correlator} for notation), by Fourier transforming 
the hadronic matrix elements of quark-quark operators
\begin{equation}
\Phi_{i \Lambda', \;j \Lambda }(k,k',P,P^{\,\prime}) =
\frac{1}{(2\pi)^4}\int d^{\,4}z\;e^{i\,(k+k')\cdot {z}/{2}} \;
\me{P^{\,\prime},\Lambda'}{\overline\psi_i(-\frac{z}{2})\,
\psi_j(\frac{z}{2})}{P,\Lambda} \eqpt
\label{eqncorrelazioneelicita}
\end{equation}
We assume the hadron is a spin $\frac{1}{2}$ particle, say a nucleonic target, which is in an eigenstate of 
light-cone helicity characterized by the initial and final light-cone helicity, 
$\Lambda$ and $\Lambda'$ respectively, which are defined from the spin vectors $S^\mu$ and ${S^{'\mu}}$
\begin{equation}
S^\mu = \frac{\Lambda}{m}(P^\mu - \frac{m^2}{P^+}v^{'\mu})  \quad , \quad
{S'}^\mu =\frac{\Lambda'}{m}({P^{\,\prime}}^\mu - \frac{m^2}{{P^{\,\prime}}^+}
v^{'\mu})
\eqcm
\label{eqnspinori}
\end{equation}
where $m$ is the hadron mass, and $v^{'\mu}$ is the null vector on the 
light-cone $v^{'\mu}=[0,1,\vec 0_{T}]$. Throughout the paper we use the 
component notation $u=[u^+,u^-,\vec u_{T}]$ with 
$u^\pm=(u^0\pm u^3)/\sqrt{2}$ and the transverse part $\vec u_{T}=(u^1,u^2)$. 
Explicit representations of spinors for light-cone helicity eigenstates are 
given for instance by Kogut and Soper~\cite{Kogut:1970xa}~or 
Brodsky and Lepage~\cite{Brodsky:1989pv}.
The non-forward correlation function 
in Eq.~(\ref{eqncorrelazioneelicita}) represents a transition matrix 
element of quark-quark operators and not an expectation value like the 
forward quark-quark correlation function~\cite{Soper:1977jc}. Therefore 
it provides more general information about hadrons compared to the ordinary 
quark-quark correlation function.

The known properties of the quark-quark correlation function 
determine the structure of an ansatz, which will be the starting 
point for our analysis. The quark-quark correlation function in 
the helicity basis, defined in Eq.~(\ref{eqncorrelazioneelicita}), 
is a $4 \times 4$ matrix in the {\em partonic} Dirac space, 
labelled by $i$ and $j$, and a $2 \times 2$ matrix in the {\em hadronic} 
helicity space, labelled by $\Lambda$ and $\Lambda'$. Therefore we formulate an ansatz given by the product of a 
partonic and a hadronic sector separately. The partonic sector is 
spanned by the set ${\hat \Gamma}_{ij}$ of the 16 independent 
$4\times 4$ partonic Dirac matrices, and the hadronic sector is 
represented by all possible independent spinorial products, 
$\bar u_k(P^{\,\prime},\Lambda') \, {\Gamma}_{kl} \, u_{l}(P,\Lambda)$, 
evaluated between final and initial light-cone helicity states. Since the correlation function in 
Eq.~(\ref{eqncorrelazioneelicita}) is a scalar in Lorentz space, 
we saturate the open indices of the tensorial product of hadronic 
and partonic sectors with all possible independent tensors, $t_{\mu_{1}\cdots \mu_{p} \nu_{1} \cdots  \nu_{h} }$, constructed from 
the kinematical variables $\bar k= (k+k')/2$, $\bar P=(P+P^{\,\prime})/2$, and $\Delta=P^{\,\prime}-P$, the metric 
tensor $g_{\alpha\beta}$ and one antisymmetric tensor $\epsilon_{\alpha\beta\rho\sigma}$, leading to the ansatz of the form
\begin{equation}
\Phi_{i\Lambda',j \Lambda}(\bar k,\bar P, \Delta) =
{\hat \Gamma_{ij}}^{\,\mu_{1} \cdots  \mu_{p}}
\;\;\;
\left[
\bar u_k(P^{\,\prime},\Lambda') \,
{\Gamma_{kl}}^{\nu_{1}\cdots\nu_{h}}\,
u_l(P,\Lambda)\right]
\;\;  \;\;
t_{\,\mu_{1}\cdots \mu_{p}\,\nu_{1} \cdots  \nu_{h} } (\bar P,\bar k, \Delta)
\eqpt
\label{eqnprodotto_an}
\end{equation}
In the {\em hadronic} sector the spinorial products (spinor indices suppressed)
\begin{equation}
\bar u(P^{\,\prime},\Lambda')\; u(P,\Lambda)
\eqcm \qquad
\bar u(P^{\,\prime},\Lambda')\; \gamma_5\;u(P,\Lambda)
\eqcm \qquad
\bar u(P^{\,\prime},\Lambda')\; \sigma^{\alpha\beta}\;u(P,\Lambda)
\eqcm
\label{spinprod-ind}
\end{equation}
\begin{equation}
\bar u(P^{\,\prime},\Lambda')\; \gamma^\alpha \;u(P,\Lambda)
\eqcm \qquad
\bar u(P^{\,\prime},\Lambda')\; \gamma^\alpha\gamma_5\;u(P,\Lambda)
\label{spinprod-dep}
\end{equation}
occur. From the Gordon identities we can deduce that only the three spinor products in (\ref{spinprod-ind}) 
are independent. These spinorial products are evaluated 
in a specific frame in Appendix \ref{sec:spinorialproducts}. Furthermore, not 
all possible contractions of indices have to be taken into account in the construction of the ansatz for 
the correlation function. The relation~\cite{Diehl:2001pm}
\begin{equation}
0 \;=\; \bar u(P^{\,\prime},\Lambda') \; u(P,\Lambda)\, \Delta^\alpha +
\bar u(P^{\,\prime},\Lambda')\; i \, \sigma^{\alpha\beta} \, 2 \bar P_\beta \,
u(P,\Lambda)
\end{equation}
shows that contractions of the tensorial spinor product 
$\bar u\,\sigma^{\alpha\beta}\,u$ with $\bar P_\beta$ reduce 
to the scalar spinor product multiplied with $\Delta^\alpha$, and the 
relation~\cite{Diehl:2001pm}
\begin{equation}
0 \;=\;
\bar u(P^{\,\prime},\Lambda') \; \g_5 \, u(P,\Lambda)\, 2 {\bar P}^\alpha
+ \bar u(P^{\,\prime},\Lambda')\;
      i \, \sigma^{\alpha\beta} \,\g_5 \, \Delta_\beta\,
       u(P,\Lambda)
\label{wahl}
\end{equation}
together with the known identities
\begin{eqnarray}
\sig^{\alpha\beta} \eps_{\alpha\beta\rho\sig} &=&
-2\,i\,\sig_{\rho\sig}\,\g_5
\nn\\
\sigma^{\alpha\beta} \, \epsilon_{\alpha\rho\sig\tau}
& = &
-i \g_5 (\sig_{\rho\sig}\, g^\beta_\tau
       - \sig_{\rho\tau}\, g^\beta_\sig
       + \sig_{\sig\tau}\, g^\beta_\rho )
\label{eq:sigeps}
\end{eqnarray}
entails
\begin{eqnarray}
\bar u(P^{\,\prime},\Lambda') \;\sigma^{\alpha\beta}\; u(P,\Lambda) \;
\eps_{\alpha\beta\rho\sig} \Delta^\rho &=&
4\, \bar u(P^{\,\prime},\Lambda') \;\g_5\; u(P,\Lambda) \; {\bar P}_\sig
\eqcm \nn\\[1ex]
\bar u(P^{\,\prime},\Lambda') \;\sigma^{\alpha\beta}\; u(P,\Lambda) \;
\eps_{\alpha\rho\sig\tau} \Delta^\rho &=&
\bar u(P^{\,\prime},\Lambda') \;\g_5\; u(P,\Lambda) \;
\left({2\bar P}_\sig g^\beta_\tau - {2\bar P}_\tau g^\beta_\sig\right)
\nn\\
&& {}+\frac12
\bar u(P^{\,\prime},\Lambda') \;\sigma^{\g\delta}\; u(P,\Lambda) \;
\eps_{\g\delta\sig\tau} \Delta^\beta
\eqcm
\end{eqnarray}
which reveals that the tensorial spinor product 
$\bar u\,\sigma^{\alpha\beta}\,u$ contracted with either 
$\eps_{\alpha\beta\rho\sig}\Delta^\rho$ or 
$\eps_{\alpha\rho\sig\tau}\Delta^\rho$ 
is proportional to structures already accounted for. Note that the tensors  
$t_{ \mu_{1}\cdots \mu_{p} \nu_{1} \cdots  \nu_{h} }(\bar P,\bar k, \Delta)$ contain at 
most one Levi-Civita symbol since tensors with more than one do not result in new structures. The product of two Levi-Civita tensors 
reduces to the product of Kronecker symbols, from which no new structure arises. In a similar way the case of tensors $t_{ \mu_{1} 
\cdots \mu_{p} \nu_{1} \cdots  \nu_{h} } (\bar P,\bar k,\Delta)$ with more than two Levi-Civita symbols can be excluded.

The correlation function (\ref{eqncorrelazioneelicita}) fulfills the following 
constraints derived from properties of 
the Dirac quark fields and of the hadronic states under 
hermitian conjugation, parity and time reversal transformations (cf.~\cite{Levelt:1993ac,Mulders:1995dh})
\begin{eqnarray*}
\Phi_{i\Lambda',j\Lambda}^\dagger (k,k',P,P^{\,\prime}) & = &
\gamma_0\,\Phi_{i\Lambda, j\Lambda'}(k',k,P^{\,\prime},P)\,\gamma_0
\qquad\quad \mbox{(hermiticity constraint)} \nn \\[2mm]
\Phi_{i\Lambda',j\Lambda}(k,k',P,P^{\,\prime}) & = &
\gamma_0\,\Phi_{i-\Lambda',j-\Lambda}(\td k,\td k',\td P,\td P^{\,\prime})
\,\gamma_0
\qquad\quad \mbox{(parity constraint)} \nn \\[2mm]
{\Phi^\ast}_{i\Lambda',j\Lambda}(k,k',P,P^{\,\prime}) & = & (-i\gamma_5 C) \,
\Phi_{i\Lambda',j\Lambda } \, (\td k,\td k',\td P,\td P^{\,\prime})\,
(-i\gamma_5 C)
\qquad \mbox{(time reversal constraint)} \eqcm
\end{eqnarray*}
\begin{equation}
\label{constraints}
\end{equation}
where the notation ${\tilde u = (u^0,-\vec{u})}$ for momenta and spin vectors is used. 
These constraints can be implemented in building the ansatz for 
the correlation function. As far as the time reversal constraint is concerned, Collins 
has shown that the time reversal constraint is not applicable if a Wilson line is inserted in the
quark correlation function. Therefore imposing the time reversal constraint corresponds to defining the correlator 
without a Wilson line.\cite{Collins:2002}

Taking into account all the above 
information leads to the most general ansatz for $\Phi$. The lengthy 
expressions are an intermediate result of our investigation and are 
explicitly given in Appendix~\ref{sec:ansatz}.

\section{Generalized parton distributions\label{sec:GPDs}}
The generalized parton distributions 
$\Phi_{i\Lambda',j\Lambda}^{[\Gamma]}(x,\xi;t)$ are obtained as traces of the 
quark-quark correlation function with the Dirac matrices $\Gamma$, 
integrated over the transverse and minus components of the 
quark momentum, ${\vec{\bar k}}_{T}$ and $\bar k$, respectively,
\begin{equation}
\Phi_{\Lambda',\Lambda}^{[\Gamma]}(x,\xi;t)
= \frac{1}{2} \, \int \, d^2
{\vec{\bar k}}_{T} \,  d \bar k^- \, \, \,
Tr[\Phi_{\Lambda',\Lambda} \, \Gamma]
\eqcm
\label{traccia2}
\end{equation}
where $x$ and $\xi$ are the light-cone momentum fractions $x\equiv\bar k^+/\bar P^+$ and 
$\xi\equiv -\Delta^+/(2\bar P^+)$ and the Mandelstam variable $t=\Delta^2$ 
denotes the momentum transfer squared. 
The projections $\Phi_{\Lambda',\Lambda}^{[\Gamma]}$ carry only 
hadron helicity indices since 
the parton Dirac indices have been saturated in taking the trace. 
Following \cite{Ralston:1979ys,Mulders:1995dh} we rewrite the integral in Eq.~(\ref{traccia2}) with 
respect to the covariant integration variables, 
$ \sigma = 2 \bar P \cdot \bar k$ and $\tau = {\bar k}^2$, and 
the azimuthal angle $\phi$
\begin{equation}
\Phi_{\Lambda',\Lambda}^{[\Gamma]}(x,\xi;t) =  \int \, d\, \sigma \,
d\,\tau \, d\,\phi \, \, \theta(x \, \sigma - x^2 \, m^2  +
\frac{x^2 \, t}{4} - \tau)
\, \frac{Tr[\Phi_{\Lambda',\Lambda} \, \Gamma]}{4 \bar P^+} \eqpt
\label{eqnintegraleoffforward}
\end{equation}
The projections of the quark-quark correlation function $\Phi$ with the various 
Dirac matrices $\Gamma$ in Eq.~(\ref{eqnintegraleoffforward})~correspond to the 
different generalized distribution functions and determine also which 
distribution functions occur in a non-forward hard process 
at different orders in $1/\bar P^+$, $\bar P^+$ scaling with 
the hard scale in the process. In particular the leading order (~twist 2~) 
distribution functions~\cite{Ralston:1979ys,Jaffe:zw,Mulders:1995dh} 
are obtained by projecting out the ansatz for the correlation function 
with the Dirac matrices $\g^+$, $\g^+\g_5$ and $i \, \sigma^{+i}\g_5$.

\subsection{Unpolarized parton distribution}
Let us consider the unpolarized generalized distribution functions of 
the proton. Thus we project the non-forward quark-quark 
operator with the matrix $\gamma^+$. 
Following \cite{Ji:1996nm} using the notation of \cite{Diehl:1998kh} the 
GPDs $H(x,\xi;t)$ and $E(x,\xi;t)$ are defined by
\begin{eqnarray}
\label{eq:quarkSPDdef}
\Phi^{[\g^+]}_{\Lambda'\Lambda} &\equiv&
\frac{1}{2} \;
\int\frac{d\,z^-}{2\pi}\;e^{i\, x\, \bar P^{\,+}z^-}\;
\langle P^{\,\prime},\Lambda'|
        \,\bar \psi(-z/2)\,\gamma^+\,\psi^\adj(z/2)\,
                                 |P,\Lambda\rangle \,|_{z^+=0,\vec{z_\sT}=0}
\nn\\[1\baselineskip]
&=&
 \frac{\bar  u(P^{\,\prime},\Lambda')\gamma^+ u(P,\Lambda)}
      {2\bar P^{\,+}}\;
H(x,\xi;t)
+\frac{\bar u(P^{\,\prime},\Lambda')
         {i\sigma^{+\alpha}\Delta_\alpha} u(P,\Lambda)}
      {4m\,\bar P^{\,+}}\;
E(x,\xi;t) \eqpt
\nn\\[1\baselineskip]
\end{eqnarray}
Evaluating the quantity ${\Phi}^{[\g^+]}_{\Lambda'\Lambda}$ for both 
proton helicity flip ($\Lambda' = - \Lambda$) and helicity non-flip~ 
($\Lambda' = \Lambda$), the generalized distribution functions, 
$H$ and $E$, result from the following set of two equations
\begin{eqnarray}
\label{eq:quark-helcomb}
{\Phi_{++}^{[\g^+]}}= \phantom{-( } {\Phi_{--}^{[\g^+]}}
\phantom{ )^*} &=&
      {\sqrt{1-\xi^2}} \, \, \, \, \, (H - \frac{\xi^2}{1-\xi^2}\, E) \eqcm
\nn\\
{\Phi_{-+}^{[\g^+]}} = {-\left(\Phi_{+-}^{[\g^+]}\right)^*}  &=&
     \eta \, \, \frac{\sqrt{t_0-t}}{2m} \, \, E
\end{eqnarray}
with $t_0$ defined as
\begin{equation}
- t_0 = \frac{4 \xi^2 m^2}{1- \xi^2} \eqcm
\end{equation}
and a phase $\eta$ given by
\begin{equation}
\eta = \frac{\Delta^1 + i \Delta^2}{| {\bf \Delta}_\perp | }
\eqpt
\label{eqnbareta-1}
\end{equation}
Substituting the expression of the ansatz (\ref{ansatzfinalmente}) and (\ref{ansatzfinalmenteoff}) 
for the helicity non-flip and helicity flip correlation functions, 
$\Phi_{++}$ or $\Phi_{--}$, and $\Phi_{-+}$ or $\Phi_{+-}$, respectively, in 
(\ref{eqnintegraleoffforward}) and tracing with 
$\g^+$ we have a set of two equations for the two 
unknown distribution functions $H$ and $E$, which is solvable 
and gives the two unpolarized generalized parton distribution functions as
\begin{eqnarray}
H & = & \frac{1}{\sqrt{1-\xi^2}} \,\left(
    A^{(1)}(x,\xi;t)
  - \frac{ 2m \, \xi^2 }{\sqrt{1-\xi^2} \,\sqrt{t_0-t} } \,
    A^{(3)} (x,\xi;t)
\right) \nn \\
E & = & - \frac{2 \, m }{\sqrt{t_0-t}} \,A^{(3)} (x,\xi;t)
\eqcm
\nn \\
\label{equdistr}
\end{eqnarray}
where we have introduced the function $A^{(1)}(x,\xi;t)$ and 
$A^{(3)}(x,\xi;t)$ defined through the coefficients $d^{(\kappa)}_m$ 
of Eqs.~(\ref{ansatzfinalmente}) and (\ref{ansatzfinalmenteoff})
\begin{equation}
A^{(1)} (x,\xi;t) =
 \frac{1}{\sqrt{1-{\xi}^2}} \, \int \, d\, \sigma \, d\,\tau \, d\,\phi \, \, \theta(x \, \sigma - x^2 \, m^2  +
\frac{x^2 \, t}{4} - \tau) \, [d^{(1)}_{5} + x \, d^{(1)}_{7}  + 4 \xi^2 \,d^{(1)}_{10} ]
\end{equation}
and
\begin{eqnarray}
A^{(3)} (x,\xi;t) & = &
 \int \, d\, \sigma \, d\,\tau \, d\,\phi \, \, 
\theta(x \, \sigma - x^2 \, m^2  
+ \frac{x^2 \, t}{4} - \tau) \, \, \eta \nn \\
&& 
\left. \{ \frac{m}{\sqrt{t-t_0} \, \sqrt{1-{\xi}^2}} \, (-2\, \xi) \,
[d^{(3)}_{6} + x \,d^{(3)}_{9} + d^{(3)}_{11} + 4\, {\xi}^2 \,d^{(3)}_{13} ] \right. \nn \\
&& 
\left.
-2\, \xi \,\frac{\sqrt{t_0-t}}{m} \, d^{(3)}_{112}
-2\, \xi \,\frac{m}{\sqrt{t-t_0}} \, d^{(3)}_{96} \right \}
\eqpt\end{eqnarray}
The expressions for the generalized distribution functions $H$ and $E$ in Eq.~(\ref{equdistr}) do 
not contain any term in the amplitudes $d^{(4)}_m$ and $d^{(2)}_m$, since these amplitudes are 
related to the amplitudes $d^{(1)}_m$ and $d^{(3)}_m$, respectively, as shown in Eqs.~(\ref{relazionid1d4_parity}), 
(\ref{relazionid1d4_hermiticity}), (\ref{relazionid1d4_timereversal}), (\ref{relazionid2d3_parity}), (\ref{relazionid2d3_hermiticity})
and (\ref{relazionid2d3_timereversal}).

In the forward case rotational invariance around the collinear axis implies 
the conservation of the longitudinal component of 
angular momentum, i.e. it requires total helicity to be conserved 
(refer to Fig.\ref{fig:correlator})
\begin{equation}
\Lambda + \lambda' = \Lambda' + \lambda \; .
\label{eqnhelicityconservation}
\end{equation}
Helicity conservation in Eq.~(\ref{eqnhelicityconservation}) then shows a 
link between the quark and nucleon helicity degrees of freedom. Through the 
projection of the quark-quark correlation function with the matrix 
$\gamma^+$ the nucleon helicity does not flip and the only possible 
hadron helicity combinations in the forward limit are $+ \,+$ and $-\,-$. 

\subsection{Polarized parton distributions}
The polarized quark distributions, 
$\widetilde{H}(x,\xi;t)$ and $\widetilde{E}(x,\xi;t)$, 
are defined by the Fourier transforms of the axial vector matrix element
\begin{eqnarray}
\Phi_{\Lambda'\Lambda}^{[\g^+\g_5]} &\equiv&
\frac{1}{2} \;
\int\frac{d\,z^-}{2\pi}\;e^{i\,x\,\bar P^{\,+}z^-}\;
\langle P^{\,\prime},\Lambda'|
        \,\bar \psi_q(-\frac{z}{2})\,\gamma^+\gamma_5\,\psi_q(\frac{z}{2})\,
                       |P,\Lambda\rangle\,|_{z^+=0,\vec{z_\sT}=0}
\nn\\[1\baselineskip]
&=&
 \frac{\bar u(P^{\,\prime},\Lambda')\gamma^+\gamma_5 u(P,\Lambda)}
      {2\bar P^{\,+}}\;
\widetilde{H}(x,\xi;t)
+\frac{\bar u(P^{\,\prime},\Lambda')\Delta^+\gamma_5 u(P,\Lambda)}
      {4m\,\bar P^{\,+}}\;
\widetilde{E}(x,\xi;t)
\eqpt
\end{eqnarray}
For the different proton helicity combinations we now find
\begin{equation}
\label{eq:pol-quark-helcomb}
\Phi_{++}^{[\g^+\g_5]} =
-\Phi_{--}^{[\g^+\g_5]}  = {\sqrt{1-\xi^2}} \, \, \, \,
(\widetilde{H} - \frac{\xi^2}{1-\xi^2}\, \widetilde{E})
\eqcm
\end{equation}
and
\begin{equation}
\Phi_{-+}^{[\g^+\g_5]}=
\left(\Phi_{-+}^{[\g^+\g_5]}\right)^* =
\eta \, \frac{\sqrt{t_0 -t}}{2m} \, \xi \, \tilde E
\eqpt
\label{eqnmavagiu'}
\end{equation}
Substituting the ans\"atze (\ref{ansatzfinalmente}) and (\ref{ansatzfinalmenteoff}) 
for the correlation functions we obtain a set of two equations in the two unknown functions $\tilde H$ and $\tilde E$
\begin{eqnarray}
\tilde H & = & \frac{1}{\sqrt{1-\xi^2}} \,
\left(  B^{(1)} (x,\xi;t)
- \frac{ 2m \,\xi }{\sqrt{1-\xi^2} \,\sqrt{t_0-t} } \,
B^{(3)} (x,\xi;t) \right)
\nn \\
\tilde E & = & - \frac{2 \, m  }{\xi \, \sqrt{t_0-t}} \,
B^{(3)}(x,\xi;t)
\eqcm
\end{eqnarray}
where we have introduced the function $B^{(1)}(x,\xi;t)$ and 
$B^{(3)}(x,\xi;t)$ defined as\\
\begin{eqnarray*}
B^{(1)} (x,\xi;t) & = &    \frac{1}{\sqrt{1-{\xi}^2}} \, \int \, d\, \sigma \,
d\,\tau \, d\,\phi \, \, \theta(x \, \sigma - x^2 \, m^2  +
\frac{x^2 \, t}{4} - \tau) 
[ d^{(1)}_{12} + x \, d^{(1)}_{14} + 4 \xi^2 \, d^{(1)}_{17}]
\end{eqnarray*}
\begin{equation}
\label{eqn:b1}
\end{equation}
and
\begin{eqnarray}
B^{(3)} (x,\xi;t) & = &
 \int \, d\, \sigma \, d\,\tau \, d\,\phi \, \, 
\theta(x \, \sigma - x^2 \, m^2  + \frac{x^2 \, t}{4} - \tau)  \, \, \eta \nn \\
&& 
\left \{
\frac{m}{\sqrt{t-t_0} \, \sqrt{1-{\xi}^2}}
[d^{(3)}_{16} + 4\, {\xi}^2 \, d^{(3)}_{18} \, + x \,(d^{(3)}_{19} 
+ 4\, {\xi}^2 \,d^{(3)}_{21} ) + 4\, {\xi}^2 \,d^{(3)}_{23} 
 \right.\nn \\
&+&
\left. \frac{m}{{\sqrt{t-t_0}}}\, [2 \, {\xi} \, d^{(3)}_{97} - x \,d^{(3)}_{98}] - \frac{\sqrt{t_0-t}}{m} \, [d^{(3)}_{113} + x \,d^{(3)}_{114}] \right \}
\eqpt
\end{eqnarray}
respectively. 

\subsection{Parton helicity changing distributions}
There are also twist 2 generalized distributions that 
change the helicity of the active parton. The 
corresponding quark distributions are constructed from the operator 
$\bar \psi_q\,i\,\sigma^{+i}\gamma_5\,\psi_q$. By counting the helicity amplitudes 
Hoodbhoy and Ji~\cite{Hoodbhoy:1998vm} introduce two independent 
quark helicity changing generalized distributions corresponding to hadron 
helicity flip and non-flip terms. However, Diehl~\cite{Diehl:2001pm} claims 
that there are four 
independent parton helicity changing generalized distributions defined by
\begin{eqnarray*}
{{\cal G}^i_{\Lambda'\Lambda}} &\equiv&
\frac{1}{2\sqrt{1-\xi^2}} \;
\int\frac{d\,z^-}{2\pi}\;e^{i\, x\, \bar P^{\,+}z^-}\;
\langle P^{\,\prime},\Lambda'|
   \,\bar \psi_q(-\frac{z}{2})\,i\,\sigma^{+i}\g_5 \,\psi_q^\adj(\frac{z}{2})\,
                           |P,\Lambda\rangle\,|_{z^+=0,{\bf z_\sT}=0}
\nn\\[1\baselineskip]
&=&
\bar  u(P^{\,\prime},\Lambda')
\,i\,\sigma^{+i}\g_5\,
u(P,\Lambda)\;
H_T(x,\xi;t)
{}+\bar u(P^{\,\prime},\Lambda')
\,\frac{i\,\epsilon^{+i\alpha\beta}\Delta_\alpha\bar P_\beta}{m^2}\,
u(P,\Lambda)\;
\widetilde H_T (x,\xi;t)
\nn\\[1\baselineskip]
& + &
{}\bar  u(P^{\,\prime},\Lambda')
\,\frac{i\,\epsilon^{+i\alpha\beta}\Delta_\alpha\g_\beta}{2m}\,
u(P,\Lambda) \;
E_T(x,\xi;t)
{}+\bar u(P^{\,\prime},\Lambda')
\,\frac{i\,\epsilon^{+i\alpha\beta}\bar P_\alpha\g_\beta}{m}\,
u(P,\Lambda) \;
\widetilde E_T (x,\xi;t)
\eqpt
\end{eqnarray*}
\begin{equation}
\label{markus}
\end{equation}
In Eq.~(\ref{markus}) the Lorentz index $i$ takes the values {1,2}). 
In the following we show that the same conclusion concerning 
the number of independent helicity changing GPDs is obtained by tracing the 
ansatz for the quark-quark correlation given in Eq.~(\ref{ansatzfinalmente}) 
with the matrix $\psi_q\,i\,\sigma^{+i}\gamma_5$. Following Diehl~\cite{Diehl:2001pm} we define
\be
{{\cal G}^i_{\Lambda'\Lambda}}= \frac{\Phi_{\Lambda'\Lambda}^
{[i \sig^{i+}\g_5]}}{\sqrt{1-\xi^2}} \,.
\label{eqnDiehl:2001pm}
\ee
By comparing the definitions for ${{\cal G}^i}_{\Lambda'\Lambda}$ 
in Eq.~(\ref{eqnDiehl:2001pm}) and 
the projections of the non-forward correlation function $\Phi$ with matrix 
$i \,\sigma^{i+}\g_5$, the GPDs, $H_T$, $\tilde H_T$, $E_T$, and 
$\tilde E_T$, arise. In fact the trace of the ansatz for 
the non-forward quark-quark correlation 
function with the matrix $i \,\sigma^{i+}\g_5$ gives a system of four linear 
independent equations
\begin{eqnarray}
\Phi_{++}^{[i \,\sigma^{i+}\g_5]}(x,\xi;t) &=& {C^{(1)}}^i + {D^{(1)}}^i \nn \\
\Phi_{--}^{[i \,\sigma^{i+}\g_5]}(x,\xi;t) &=& {C^{(1)}}^i - {D^{(1)}}^i \nn \\
\Phi_{-+}^{[i \,\sigma^{i+}\g_5]}(x,\xi;t) &=& {C^{(3)}}^i + {D^{(3)}}^i \nn \\
\Phi_{+-}^{[i \,\sigma^{i+}\g_5]}(x,\xi;t) &=& {{C^{(3)}}^i}^* - {{D^{(3)}}^i}^*
\label{sistemagiusto}
\end{eqnarray}
where ${C^{(1)}}^i$, expressed in terms of the coefficients of the ansatz, reads
\begin{eqnarray*}
{C^{(1)}}^i &=& \frac{1}{\sqrt{1-\xi^2}} \, 
 \int \, d\, \sigma \,
d\,\tau \, d\,\phi \, \, \theta(x \, \sigma - x^2 \, m^2  +
\frac{x^2 \, t}{4} - \tau) \, i \left.
 \Bigg \{ -\frac{\eps^{+ i\rho\sig}}{m \,\bar P^+} \bar P_\rho \Delta_\sig \,
 d^{(1)}_{21} + \frac{\eps^{\rho\sigma + i}}{m \, \bar P^+} \bar k_\rho  \Delta_\sig \, d^{(1)}_{23} \right. \Bigg \}
\end{eqnarray*}
\begin{equation}
\label{skewed2}
\end{equation}
and ${D^{(1)}}^i$ reads
\begin{eqnarray}
{D^{(1)}}^i &=& \frac{1}{\sqrt{1-\xi^2}} \,
 \int \, d\, \sigma \,
d\,\tau \, d\,\phi \, \, \theta(x \, \sigma - x^2 \, m^2  +
\frac{x^2 \, t}{4} - \tau) \, \left. \Bigg \{ \Delta^i \,
\Big [ \frac{i}{m^3} \, (\xi \, \sigma - 2 \, m^2 \,\, \xi \,x + \bar k \cdot \Delta  )  \,d^{(1)}_{56}
\right. \nn \\ 
& + &
\left.
\frac{i}{m^3} \,( \xi \, x \, \sigma -  2 \, \xi \, \tau  - x \, \bar k \cdot \Delta) \, d^{(1)}_{57}
+ \frac{2\, i \, \xi}{m} \, ( d^{(1)}_{31} - x\,d^{(1)}_{33} ) \Big ] \right. \Bigg \} \,,
\label{skewed1}
\end{eqnarray}
where $\bar k \cdot \Delta = x \, (m^2 - \frac{t}{4}) \, ( \xi + 1) - \xi \, \sigma - \bar k_{T} \cdot \Delta_{T} $. 

The function ${C^{(3)}}^i$ can be written as
\begin{eqnarray*}
{C^{(3)}}^i &=&
 \int \, d\, \sigma \,
d\,\tau \, d\,\phi \, \, \theta(x \, \sigma - x^2 \, m^2  +
\frac{x^2 \, t}{4} - \tau) \, \eta   \, \, \left. \Bigg
\{\frac{m}{\sqrt{t-t_0} \, \sqrt{1-{\xi}^2}}[  \right. \nn \\
& + & 
 \left. 
\frac{\Delta^i}{m^3} \, d^{(3)}_{88}(2 \, m^2 \, \xi \,x -2  \xi \,\bar k \cdot \bar P - \bar k \cdot \Delta ) + \frac{\Delta^i}{m^3} \, 2 \, \xi \, d^{(3)}_{90}(2\, \xi\,\bar k \cdot \bar P - 2 \xi \, \,{\bar k}^2  - 
x \,\bar k \cdot \Delta) \right. \nn \\
&- 2 &\, \frac{\eps^{\rho\sigma + i}}{m \,\bar P^+} \bar P_\rho \Delta_\sig \, \xi \, d^{(3)}_{31} -  \left.
2 \, \frac{\eps^{\rho\sigma + i}}{m \,\bar P^+} \bar k_\rho \Delta_\sig \, \xi \, d^{(3)}_{34} \,  
+ \frac{\Delta^i}{m^3} ]\right.
\Bigg \}
\end{eqnarray*}
\begin{equation}
\label{skewed3}
\end{equation}
and ${D^{(3)}}^i$ in terms of the amplitudes $d^{(3)}_{m}$ is
\begin{eqnarray*}
{D^{(3)}}^i &=&
 \int \, d\, \sigma \,
d\,\tau \, d\,\phi \, \, \theta(x \, \sigma - x^2 \, m^2  +
\frac{x^2 \, t}{4} - \tau) \, \eta   \, \left. \Bigg
\{ \frac{m}{\sqrt{t-t_0} \, \sqrt{1-{\xi}^2}} [
\frac{\Delta^i}{m}2\, i \, \xi \, (d^{(3)}_{46}+x\,d^{(3)}_{49})   \right. \nn \\
&+&  \left. \frac{\Delta^i}{m^3} \, 2 \, \xi \, d^{(3)}_{86}( x^2 \, m^2 - 2 \,x\,
\bar k \cdot \bar P + {\bar k}^2 ) ]\right. \Bigg \}
\end{eqnarray*}
\begin{equation}
\label{skewed4}
\end{equation}
Inserting in Eq.~(\ref{markus}) the results for the spinorial products, given in the Appendix \ref{sec:spinorialproducts}, we obtain 
the four possible helicity combinations
\begin{eqnarray}
{{\cal G}^j_{++}} & = &
  H_T \, \frac{\eps^{+j\rho\sigma}\bar P_\rho \Delta_\sig \, \xi \,
  \sqrt{1-\xi^2} }{(1+ \xi) \, m} \nn \\
&& - i \, \widetilde{H_T} \,2m \,\eps^{+j\rho\sigma}\bar P_\rho \Delta_\sig \,
  \sqrt{1-\xi^2} \nn \\
&&{}+ E_T \, \frac{2 \,(\bar P^+ \Delta^j  \ \xi + i \,
  \eps^{+j\rho\sigma}\bar P_\rho \Delta_\sig ) }
  {\sqrt{1-\xi^2}}  \nn \\
&&{}+ \widetilde{E_T} \,  \frac{- i\,\bar P^+ \,\Delta^j + \xi \,
  \eps^{+j\rho\sigma}\bar P_\rho \Delta_\sig }
  {\sqrt{1-\xi^2}}
\label{eqn:G++}
\end{eqnarray}
\begin{eqnarray}
{{\cal G}^j_{--}} & = &
  H_T \,  \frac{\eps^{+j\rho\sigma}\bar P_\rho \Delta_\sig \, \xi \,
  \sqrt{1-\xi^2} }{(1+ \xi) \, m} \nn \\
&&{}- i \, \widetilde{H_T} \,2m \,\eps^{+j\rho\sigma}\bar P_\rho \Delta_\sig \,
  \sqrt{1-\xi^2} \nn \\
&&{}+ E_T \, \frac{2 \,(-  \bar P^+ \Delta^j  \, \xi + i \,
\eps^{+j\rho\sigma}\bar P_\rho \Delta_\sig ) }
  {\sqrt{1-\xi^2}} \nn \\
&&{}+ \widetilde{E_T} \,  \frac{ i\,\bar P^+ \, \Delta^j + \xi \,
  \eps^{+j\rho\sigma}\bar P_\rho \Delta_\sig }
  {\sqrt{1-\xi^2}}
\label{eqn:G--}
\end{eqnarray}
\begin{eqnarray}
{{\cal G}^j_{-+}} & = & - \eta \,  \Bigg [
   - 2 \, H_T \,  \, \frac{ - i \bar P^+ \Delta^j +
   \eps^{+j\rho\sigma}\bar P_\rho \Delta_\sig  }
  {\sqrt{t_0-t}} \nn \\
&&{}- \widetilde{H_T}\, \frac{\eps^{+j\rho\sigma}\bar P_\rho \Delta_\sig \,
  ( 4m^2 \xi^2 - \xi^2 \, t + t)  }
  {\sqrt{1-\xi^2}\sqrt{t_0-t} } \nn \\
&&{}+  4 m \, E_T \, \xi^2 \, \frac{  - \bar P^+ \Delta^j  + i \,
  \eps^{+j\rho\sigma} \bar P _\rho \Delta_\sig \,
   } { \sqrt{1-\xi^2} \,\sqrt{t_0-t} }  \nn \\
&&{}- 2m \, \xi \, \widetilde{E_T} \,  \frac{ (- \bar P^+ \, \Delta^j  + i \,
  \eps^{+j\rho\sigma}\bar P_\rho \Delta_\sig ) }
  {\sqrt{1-\xi^2} \sqrt{t_0 - t} } \Bigg ]
\label{eqn:G-+}
\end{eqnarray}
\begin{eqnarray}
{{\cal G}^j_{+-}} & = &  {\eta}^* \, \Bigg [
   -2 \, H_T \,\frac{ i \bar P^+  \Delta^j +
   \eps^{+j\rho\sigma}\bar P_\rho \Delta_\sig  }
  {\sqrt{t_0-t}} \nn \\
&&{}- \widetilde{H_T} \frac{\eps^{+j\rho\sigma}\bar P_\rho \Delta_\sig \,
  ( 4m^2 \xi^2 - \xi^2 \, t + t)  }
  {\sqrt{1-\xi^2}\sqrt{t_0-t} }  \nn \\
&&{}+  4 m \xi^2 \, E_T \, \frac{ \bar P^+ \Delta^j  + i \,
  \eps^{+j\rho\sigma} \bar P _\rho \Delta_\sig \,
   } { \sqrt{1-\xi^2} \,\sqrt{t_0-t} } \nn \\
&&{}-  2m \xi \, \widetilde{E_T} \, \frac{ \bar P^+ \, \Delta^j + i \,
  \eps^{+j\rho\sigma}\bar P_\rho \Delta_\sig  }
  {\sqrt{1-\xi^2} \sqrt{t_0 - t} } \Bigg ]
\label{eqn:G+-}
\eqpt
\end{eqnarray}
Since we are only interested in the number of independent GPDs we refrain 
from isolating $H_T$, $\tilde H_T$, $E_T$, and 
$\tilde E_T$ in Eqs.~(\ref{eqn:G++}),~(\ref{eqn:G--}),~(\ref{eqn:G-+}) and 
(\ref{eqn:G+-}). Instead, from the fact that Eq.~(\ref{sistemagiusto}) 
constitutes a set of four linearly independent equations, we conclude that 
one can define four independent GPDs from it. Thus, we have shown that the 
number of independent helicity changing generalized parton distributions 
is four, as claimed by Diehl \cite{Diehl:2001pm}.

\section{Conclusions\label{sec:conclusions} }

We have presented a detailed analysis of the non-forward 
quark-quark correlation function. Constraints on the correlation function, 
not yet in the literature, were obtained by implementing the known properties 
of the fundamental fields of QCD, quarks and gluons, under parity and time 
reversal transformations and applying hermiticity. We developed a new method to construct an ansatz 
for the correlation function. The quark-quark correlation function could then be expressed in terms of tensorial structures 
formed by the independent dynamical vectors and by Dirac matrices. The constraints obtained were implemented to reduce the number of 
independent amplitudes multiplying these tensorial 
structures in the ansatz.

Finally we projected out the leading order GPDs, i.e. 
we expressed the unpolarized, polarized and 
parton helicity flip distributions in terms of 
the amplitudes occurring in the ansatz. The formalism adopted allowed us to 
conclude that the number of independent parton helicity changing distributions 
is four, in agreement with Diehl's argument~\cite{Diehl:2001pm}. 
We stress that the result about the number of the independent 
GPDs was obtained by Diehl in a completely different way and this 
is a confirmation of both methods used to approach the problem. 
On one hand we wrote the 
most general ansatz which can describe non-forward quark-quark 
correlation functions, i.e. we represented matrix elements of 
non-local non-forward quark-quark operators in terms of tensorial structures, 
built from the involved momenta on the basis of general properties of 
invariance. Then we traced the ansatz for the non-forward correlation function 
with different Dirac matrices and we 
could read off which of these structures contribute to each GPD. On 
the other hand Diehl's approach was to count 
the number of independent helicity amplitudes occurring in DVCS cross 
sections on the basis of time reversal and parity invariance which these 
amplitudes have to fulfill.

The advantage of having built an ansatz for the non-forward 
quark-quark correlation function is that, by tracing it with the 
different $\Gamma$ Dirac matrices, we gain the different generalized 
distribution functions in terms of some of the amplitudes occurring in 
the ansatz, and we are thus able to predict the dependence of GPDs 
upon the different fundamental structures entering the ansatz.

Note that in Eq.~(\ref{traccia2}) we integrate over 
$ d^2\, {\bf \vec{\bar k}}_{T} \, d\,\bar k^- $ and thus we 
consider only distribution functions which do not 
depend on the transverse momentum of the quarks ${\vec{\bar k}}_{T}$. 
We remark that, in principle by having an ansatz for the 
off-forward quark-quark correlator, 
one could extract generalized profile functions 
which depend additionally on the transverse momentum ${\bf \vec{\bar k}}_{T}$ of quarks. For instance the investigation of 
${\bf \vec{k}}_{T}$-depending ordinary parton 
distributions has been extensively carried out by many 
groups theoretically and experimental investigations are currently 
under way. For non-forward processes no formalism for the systematic study 
of ${\bf \vec{\bar {k}}}_{T}$-dependence has yet been attempted 
and an experimental program on ${\bf \vec{\bar {k}}}_{T}$ effects seems 
far beyond present abilities. In the foreseeable future there are good 
prospects to acquire some knowledge on GPDs depending on $(x,\xi,t;Q^2)$, as 
well as possibly additional ${\bf \vec{\bar {k}}}_{T}$-dependence.

We have worked out a powerful method of analysis which in the present paper 
was applied completely to the leading twist level. The same method can be 
implemented to investigate twist 3 and twist 4 generalized distribution 
functions. For instance one could expect that useful relations between 
leading and next to leading order generalized distributions could emerge as 
suggested by similar experience in the forward case. In this sense the 
present work represents a valuable starting point for further investigations.

\appendix

\section{Spinorial products\label{sec:spinorialproducts}}

The independent spinorial products of ansatz~(\ref{eqnprodotto_an}) 
listed in Eq.~(\ref{spinprod-ind}) are most easily evaluated by 
using explicit expressions for light-cone helicity spinors~\cite{Brodsky:1989pv}
\begin{equation}
u_{LC}(P,+) =
\frac{1}{(2\sqrt{2}\,P^+)^{1/2}}\left(
\begin{array}{c}
  \sqrt{2}\,P^+ + m\\
  P^1+i\,P^2\\
  \sqrt{2}\,P^+ - m\\
  P^1+i\,P^2
\end{array}\right)
\eqcm \qquad
u_{LC}(P,-) =
\frac{1}{(2\sqrt{2}\,P^+)^{1/2}}\left(
\begin{array}{c}
  -P^1+i\,P^2\\
  \sqrt{2}\,P^++m\\
  P^1-i\,P^2\\
  -\sqrt{2}\,P^+ + m
\end{array}\right)
\eqpt
\label{helicityeigenstates}
\end{equation}
normalized according to
\begin{eqnarray}
  \bar u(P,+)\,u(P,+)&=&\bar u(P,-)\,u(P,-)=2\,m
  \nn\\
  \bar u(P,-)\,u(P,+)&=&\bar u(P,+)\,u(P,-)=0
\eqpt
\label{normalizzazione}
\end{eqnarray}
In a frame of reference where the longitudinal direction is defined 
by the proton average momentum $\bar P$, the momenta of incoming and 
outgoing hadrons are parameterized as (in the light-cone component notation 
$z^\mu=[z^+,z^-,\vec{z}_\perp]$, where $z^\pm=(z^0\pm z^3)/\sqrt{2}$ 
and $\vec{z}_\perp$ is a two-dimensional transverse vector)
\begin{eqnarray}
P^\mu &=& (\bar P-\Delta/2)^\mu \;=\;
\left[(1+\xi)\bar P^+,
      \frac{m^2+\vec{\Delta}_\perp^2/4}{2(1+\xi)\bar P^+},
      -\frac{\vec{\Delta}_\perp}{2}\right]
\nn\\
P^{\,\prime\mu} &=& (\bar P+\Delta/2)^\mu \;=\;
\left[(1-\xi)\bar P^+,
      \frac{m^2+\vec{\Delta}_\perp^2/4}{2(1-\xi)\bar P^+},
      +\frac{\vec{\Delta}_\perp}{2}\right]
\end{eqnarray}
With this parametrization one obtains for the spinorial products 
in the helicity non-flip case, i.e. $\Lambda'=\Lambda=\pm 1$
\begin{eqnarray}
\bar u(P^{\,\prime},\Lambda')\;u(P,\Lambda) &=&
\frac{2\,m}{\sqrt{1-\xi^2}}
\nn\\
\bar u(P^{\,\prime},\Lambda')\;\g_5\;u(P,\Lambda) &=&
\Lambda\,\frac{2\,m\,\xi}{\sqrt{1-\xi^2}}
\nn\\
\bar u(P^{\,\prime},\Lambda')\;\sig^{\mu\nu}\;u(P,\Lambda) &=&
  {}\frac{i\,m\,\left(\Delta^\mu v^{\prime\nu}
      -\Delta^\nu v^{\prime\mu}\right)}
  {\bar P^+\,\sqrt{1-\xi^2}}
  +\Lambda\,\frac{2m\,\eps^{\mu\nu\rho\sig} \bar P_\rho v'_\sig}
  {\bar P^{\,+}\sqrt{1-\xi^2}} \eqpt
\label{eqn:prodotti1}
\end{eqnarray}
The spinorial products if the helicity is flipped and $\Lambda'=-\Lambda= 1$ are
\begin{eqnarray}
\bar u(P^{\,\prime},\Lambda')\;u(P,\Lambda)
&=&
- \eta \, \sqrt{t_0-t}
\nn\\
\bar u(P^{\,\prime},\Lambda')\;\g_5\;u(P,\Lambda)
&=&
- \eta \, \Lambda\sqrt{t_0-t}
\nn\\
\bar u(P^{\,\prime},\Lambda')\;\sig^{\mu\nu}\;u(P,\Lambda)
&=&
  - \eta \, \Big[ {}-\frac{2\,i \,\left(\bar P^\mu\Delta^\nu-\bar P^\nu\Delta^\mu\right)}
  {\sqrt{t_0-t}}
  {}-\frac{2\,i \, m^2\,
    \left(\Delta^\mu v^{\prime\nu}-\Delta^\nu v^{\prime\mu}\right)}
  {\bar P^+\,(1-\xi^2)\sqrt{t_0-t}}
  \nn\\[\bls] &&
  {}-\frac{4 \,\Lambda\, m^2\xi\,\eps^{\mu\nu\rho\sig} \bar P_\rho v'_\sig}
  {\bar P^+\,(1-\xi^2)\sqrt{t_0-t}}
  {}+\frac{2\Lambda\,\,\eps^{\mu\nu\rho\sig} \bar P_\rho \Delta_\sig}
  {\sqrt{t_0-t}} \Big ] \, ,
\label{eqn:prodotti2}
\end{eqnarray}
while if $\Lambda'=-\Lambda= -1$
\begin{eqnarray}
\bar u(P^{\,\prime},\Lambda')\;u(P,\Lambda)
&=&
{\eta}^* \, \sqrt{t_0-t}
\nn\\
\bar u(P^{\,\prime},\Lambda')\;\g_5\;u(P,\Lambda)
&=&
{\eta}^*  \, \Lambda\sqrt{t_0-t}
\nn\\
\bar u(P^{\,\prime},\Lambda')\;\sig^{\mu\nu}\;u(P,\Lambda)
&=&
{\eta}^*  \, \Big [{}-\frac{2\,i \,\left(\bar P^\mu\Delta^\nu-\bar P^\nu\Delta^\mu\right)}
  {\sqrt{t_0-t}}
  {}-\frac{2\,i \, m^2\,
    \left(\Delta^\mu v^{\prime\nu}-\Delta^\nu v^{\prime\mu}\right)}
  {\bar P^+\,(1-\xi^2)\sqrt{t_0-t}}
  \nn\\[\bls] &&
  {}-\frac{4 \,\Lambda\, m^2\xi\,\eps^{\mu\nu\rho\sig} \bar P_\rho v'_\sig}
  {\bar P^+\,(1-\xi^2)\sqrt{t_0-t}}
  {}+\frac{2\Lambda\,\,\eps^{\mu\nu\rho\sig} \bar P_\rho \Delta_\sig}
  {\sqrt{t_0-t}} \Big ]
\label{eqn:prodotti3}
\end{eqnarray}
where the phase factor is given as
\begin{equation}
\eta=\frac{\Delta_1+i\,\Delta_2}{|\vec\Delta_\perp|}
\eqcm
\label{eq:phasefactor}
\end{equation}
and
\begin{equation}
|\vec\Delta_\perp|=
\sqrt{\frac{-4\xi^2m^2}{1-\xi^2}
      +\frac{4\xi^2m^2+\vec\Delta_\perp^2}{1-\xi^2}}\,\sqrt{1-\xi^2}=
\sqrt{t_0-t}\;\sqrt{1-\xi^2}
\eqcm
\end{equation}
which contains the implicit definition of the quantity $t_0$. 
Studying the form factor decomposition of the tensor current of the 
proton \cite{Diehl:2001pm}, we need additionally the following 
Dirac bilinears. 
For the helicity non-flip case, i. e., $\Lambda'=\Lambda=\pm 1$, we have
\begin{eqnarray}
\bar u(P^{\,\prime},\Lambda')\; \eps^{+j\rho\sigma} {\bar P}_\rho
  \g_\sigma \; u(P,\Lambda)
& = &
\frac{- i\,\bar P^+ \, \Lambda \Delta^j + \xi \,
  \eps^{+j\rho\sigma}\bar P_\rho \Delta_\sig }
  {\sqrt{1-\xi^2}}
\nn\\[2mm]
\bar u(P^{\,\prime},\Lambda')\;\sig^{+j}\, \g^5 \;u(P,\Lambda)
& = &
 \frac{\eps^{+j\rho\sigma}\bar P_\rho \Delta_\sig \, \xi \,
  \sqrt{1-\xi^2} }{(1+ \xi) \, m}
\nn\\[2mm]
  \bar u(P^{\,\prime},\Lambda')\;
  \eps^{+j\rho\sigma} {\bar P}_\rho {\Delta}_\sig\; u(P,\Lambda)
& = &
- 2m \, i \,\eps^{+j\rho\sigma}\bar P_\rho \Delta_\sig \,
  \sqrt{1-\xi^2}  \nn
\\[2mm]
\bar u(P^{\,\prime},\Lambda')\;
   \eps^{+j\rho\sigma} \Delta_\rho \g_\sig \; u(P,\Lambda)
& = & \frac{2 \,( \Lambda \,\bar P^+ \Delta^j \, \xi
 + i \, \eps^{+j\rho\sigma}\bar P_\rho \Delta_\sig ) }
  {\sqrt{1-\xi^2}}
\end{eqnarray}
The results for the helicity flip case with $\Lambda'=-\Lambda= + 1$ are
 \begin{eqnarray}
\bar u(P^{\,\prime},\Lambda')\;\eps^{+j\rho\sigma}{\bar P}_\rho\g_\sigma\;
        u(P,\Lambda)
& = &
-\eta \,  \Big[ {}  - 2 \, \frac{m^2 \xi (\bar P^+ \, \Delta^j \Lambda + i \,
  \eps^{+j\rho\sigma}\bar P_\rho \Delta_\sig ) }
  {\sqrt{1-\xi^2} \sqrt{t_0 - t} } \Big] \nn \\[2mm]
\bar u(P^{\,\prime},\Lambda')\;\sig^{+j}\, \g^5 \;u(P,\Lambda) & = &
-\eta \,  \Big[ {}  -2 \, \frac{ i \bar P^+ \Lambda \Delta^j +
   \eps^{+j\rho\sigma}\bar P_\rho \Delta_\sig  }
  {\sqrt{t_0-t}}  \Big]  \nn \\[2mm]
\bar u(P^{\,\prime},\Lambda')\; \eps^{+j\rho\sigma} \bar P_\rho
   \Delta_\sig
  \;u(P,\Lambda)   & = &
-\eta \,  \Big[ {}- \frac{\eps^{+j\rho\sigma}\bar P_\rho \Delta_\sig \,
  ( 4m^2 \xi^2 - \xi^2 \, t + t)  }
  {\sqrt{1-\xi^2}\sqrt{t_0-t} }  \Big]  \nn \\[2mm]
\bar u(P^{\,\prime},\Lambda')\;
 \eps^{+j\rho\sigma}\Delta_\rho\g_\sig\;u(P,\Lambda)
&=& -\eta \,  \Big[ {}4 m \xi^2 \, \frac{\bar P^+ \Lambda \Delta^j + i \,
  \eps^{+j\rho\sigma} \bar P _\rho \Delta_\sig \,
   } { \sqrt{1-\xi^2} \,\sqrt{t_0-t} } \Big]
\eqcm
   \end{eqnarray}
and for $\Lambda'=-\Lambda= - 1$
 \begin{eqnarray}
\bar u(P^{\,\prime},\Lambda')\;\eps^{+j\rho\sigma}{\bar P}_\rho\g_\sigma\;
        u(P,\Lambda)
& = &
{\eta}^* \,  \Big[{}  - 2 \, \frac{m^2 \xi (\bar P^+ \, \Delta^j \Lambda + i \,
  \eps^{+j\rho\sigma}\bar P_\rho \Delta_\sig ) }
  {\sqrt{1-\xi^2} \sqrt{t_0 - t} } \Big ] \nn \\[2mm]
\bar u(P^{\,\prime},\Lambda')\;\sig^{+j}\, \g^5 \;u(P,\Lambda) & = &
 {\eta}^* \,  \Big[{} -2 \, \frac{ i \bar P^+ \Lambda \Delta^j +
   \eps^{+j\rho\sigma}\bar P_\rho \Delta_\sig  }
  {\sqrt{t_0-t}} \Big ] \nn \\[2mm]
\bar u(P^{\,\prime},\Lambda')\; \eps^{+j\rho\sigma} \bar P_\rho
   \Delta_\sig
  \;u(P,\Lambda)   & = &
{\eta}^* \,  \Big[{}- \frac{\eps^{+j\rho\sigma}\bar P_\rho \Delta_\sig \,
  ( 4m^2 \xi^2 - \xi^2 \, t + t)  }
  {\sqrt{1-\xi^2}\sqrt{t_0-t} }  \Big ] \nn \\[2mm]
\bar u(P^{\,\prime},\Lambda')\;
 \eps^{+j\rho\sigma}\Delta_\rho\g_\sig\;u(P,\Lambda)
&=& {\eta}^* \,  \Big[{}4 m \xi^2 \, \frac{\bar P^+ \Lambda \Delta^j + i \,
  \eps^{+j\rho\sigma} \bar P _\rho \Delta_\sig \,
   } { \sqrt{1-\xi^2} \,\sqrt{t_0-t} } \Big ]
\eqcm
   \end{eqnarray}
where the phase is again given as in Eq.~(\ref{eq:phasefactor}). Note 
that latin indices $i,j=1,2$, while greek indices run from $0$ to $3$ and 
$\eps^{+1-2}=1$, since we have adopted the convention $\epsilon^{0123}=1$.

\section{Ansatz for the non-forward quark-quark correlation function
\label{sec:ansatz}}
For the sake of convenience, in view of the numerous coefficients, 
we now switch over from explicitly 
displaying the hadron helicity indices to a more dense notation 
using the index $(\kappa)$ defined as $(1)=+\,+$, $(2)=+\,-$, 
$(3)=-\,+$, and $(4)=-\,-$ and suppressing spinor indices ${ij}$. Moreover 
contractions of momenta with Levi-Civita tensors will be written in 
the compact notation $\eps^{\alpha\beta\g\delta}\, a_\alpha b_\beta c_\g d_\delta\equiv\eps^{abcd}$.

In order to write the ansatz for the quark-quark correlation function as 
indicated in Eq.~(\ref{eqnprodotto_an}) we now multiply the independent 
spinorial products, given in Eqs.~(\ref{eqn:prodotti1}), 
(\ref{eqn:prodotti2}) and (\ref{eqn:prodotti3}), 
with the 16 independent partonic Dirac matrices and saturate the free indices 
with the following tensors (~indices $\alpha$ and $\beta$ are intended saturated 
with spinorial products, while indices $\mu$, $\nu$ with indices from partonic Dirac matrices~):
\begin{eqnarray}
&& 
1 \qquad 
\bar P^\mu \qquad 
\bar k^\mu \qquad 
\Delta^\mu \qquad 
\eps^{\mu \bar P \bar k \Delta} \qquad
\bar P^\mu \bar k^\nu \qquad  
\bar P^\mu\Delta^\nu \qquad 
\bar k^\mu\Delta^\nu \qquad 
\eps^{\mu \nu \bar P \bar k} \qquad 
\eps^{\mu \nu \bar P \Delta} \qquad 
\eps^{\mu \nu \bar k \Delta} 
\nn \\[3mm] 
&&
\bar k^\alpha \Delta^\beta \qquad
\eps^{\alpha \beta \bar P \bar k} 
\nn \\[3mm] 
&&
g_{\alpha \mu} \, \bar k^{\beta} \qquad 
g_{\alpha \mu} \, \Delta^{\beta} \qquad 
\Delta^\alpha \bar k^\beta \bar P^\mu \qquad 
\Delta^\alpha \bar k^\beta \bar k^\mu \qquad 
\Delta^\alpha \bar k^\beta \Delta^\mu \qquad 
\eps^{\mu \alpha \beta \bar P} \qquad 
\eps^{\mu \alpha \beta \bar k} 
\nn \\[3mm] 
&&
\eps^{\alpha \beta \bar P \bar k } \, \bar P^{\mu} \qquad 
\eps^{\alpha \beta \bar P \bar k } \, \bar k^{\mu} \qquad 
\eps^{\alpha \beta \bar P \bar k } \, \Delta^{\mu} \qquad 
\eps^{\alpha \mu \bar P \bar k } \, \bar k^{\beta} \qquad 
\eps^{\alpha \mu \bar P \bar k } \, \Delta^{\beta} \qquad
\eps^{\mu \bar P  \bar k \Delta} \, \bar k^\alpha  \, \Delta^\beta \qquad 
g_{\mu \alpha} \, \eps^{\beta \bar P \bar k \Delta} 
\nn \\[3mm]
&& 
g_{\mu \alpha} \, g_{\nu \beta} \qquad
g_{\mu \alpha} \, {\bar k}^\beta \, {\bar P}^\nu \qquad 
g_{\mu \alpha} \, {\bar k}^\beta \,  {\bar k}^\nu \qquad 
g_{\mu \alpha} \, {\bar k}^\beta \,  {\Delta}^\nu \qquad 
g_{\mu \alpha} \, {\Delta}^\beta \, {\bar P}^\nu \qquad 
g_{\mu \alpha} \, {\Delta}^\beta \,  {\bar k}^\nu \qquad 
g_{\mu \alpha} \, {\Delta}^\beta \,  {\Delta}^\nu 
\nn  \\[3mm]
&& 
{\bar k}^\alpha \, {\Delta}^\beta \,{\bar P}^\mu {\bar k}^\nu \qquad
{\bar k}^\alpha \, {\Delta}^\beta \,{\bar P}^\mu {\Delta}^\nu \qquad
{\bar k}^\alpha \, {\Delta}^\beta \,{\bar k}^\mu {\Delta}^\nu \qquad
\eps^{\mu \nu \alpha \beta} \qquad 
\nn \\[3mm]
&&
\eps^{\alpha \beta \bar P \bar k} \, {\bar P}^\mu \, {\bar k}^\nu \qquad 
\eps^{\alpha \beta \bar P \bar k} \, {\bar P}^\mu \, {\Delta}^\nu \qquad 
\eps^{\alpha \beta \bar P \bar k} \, {\bar k}^\mu \, {\Delta}^\nu \qquad
\nn \\[3mm]
&&
\eps^{\mu \alpha \beta \bar P } \, {\bar P}^\nu \qquad
\eps^{\mu \alpha \beta \bar P} {\bar k}^\nu  \qquad
\eps^{\mu \alpha \beta \bar P } \, {\Delta}^\nu \qquad
\eps^{\mu \alpha \beta \bar k } \, {\bar P}^\nu \qquad
\eps^{\mu \alpha \beta \bar k} {\bar k}^\nu  \qquad
\eps^{\mu \alpha \beta \bar k } \, {\Delta}^\nu
\nn \\[3mm]
&&
\eps^{\mu \nu \alpha \bar P} {\bar k}^\beta \qquad
\eps^{\mu \nu \alpha \bar P } \, {\Delta}^\beta \qquad
\eps^{\mu \nu \alpha \bar k } \, {\bar k}^\beta  \qquad
\eps^{\mu \nu \alpha \bar k } \, {\Delta}^\beta \qquad
\eps^{\mu \nu \bar P \bar k } \, {\bar k}^\alpha \, {\Delta}^\beta
\nn \\[3mm]
&&
g_{\mu \alpha}\, \eps^{\beta \nu \bar P \bar k} \qquad 
g_{\mu\alpha} \, \eps^{\nu \bar P \bar k \Delta} \bar k^\beta \qquad 
g_{\mu\alpha} \, \eps^{\nu \bar P \bar k \Delta} \Delta^\beta \qquad
\nn  \\[3mm]
&&
\eps^{\mu \nu \bar P \Delta} \, \bar k^{\alpha} \Delta^\beta \qquad
\eps^{\mu \nu \bar k \Delta} \, \bar k^{\alpha} \Delta^\beta \qquad
\eps^{\alpha \mu \bar P \bar k}  \, \bar k^{\beta} \, \bar P^\nu \qquad
\eps^{\alpha \mu \bar P \bar k}  \, \bar k^{\beta} \, \bar k^\nu 
\nn  \\[3mm]
&&
\eps^{\alpha \mu \bar P \bar k}  \, \bar k^{\beta} \, \Delta^\nu \qquad
\eps^{\alpha \mu \bar P \bar k}  \, \Delta^{\beta} \, \bar P^\nu \qquad
\eps^{\alpha \mu \bar P \bar k}  \, \Delta^{\beta} \, \bar k^\nu \qquad
\eps^{\alpha \mu \bar P \bar k}  \, \Delta^{\beta} \, \Delta^\nu 
\nn  \\[3mm]
&&
\eps^{\mu \bar P \bar k \Delta} \, \bar k^\alpha \, \Delta^\beta \, \bar P^\nu \qquad 
\eps^{\mu  \bar P \bar k \Delta} \, \bar k^\alpha \, \Delta^\beta \, \bar k^\nu \qquad
\eps^{\mu \bar P \bar k \Delta} \, \bar k^\alpha \, \Delta^\beta \, \Delta^\nu 
\end{eqnarray}
The method introduced in Eq.~(\ref{eqnprodotto_an}) produces an ansatz 
containing Lorentz tensorial structures multiplied with various 
amplitudes that are functions of all possible Lorentz scalars. In the ansatz we choose 
to indicate explicitly only the dependence on scalar 
products $\Delta\cdot v'$ involving the vector $v'$ which 
occurs in the definition of the helicity eigenstates of the hadrons. Through 
a renaming of the amplitudes each tensorial structure is thus 
multiplied by an amplitude $a^{(\kappa)}_{m}$ depending implicitly on 
all scalar products which can be formed from the momentum vectors $\bar k$, $\bar P$ and $\Delta$  
\begin{equation}
a_{m}^{(\kappa)} = a_{m}^{(\kappa)} (\bar k \cdot \bar P, \bar k^2,\bar k \cdot \Delta, t)
\eqpt
\end{equation}
For the helicity non-flipped case with $\kappa=1,4$ the ansatz reads
\begin{eqnarray*}
\Phi^{(\kappa)}(~\bar k,\bar P, \Delta~) &=& \frac{(\delta_{\kappa 1}+\delta_{\kappa 4})}{\sqrt{1-\xi^2}}
 \Bigg [m \,a^{(\kappa)}_1 
{}+  \frac{m \, \Delta \cdot v' }{\bar P^+} \,a^{(\kappa)}_{2} 
{}+ \g^5 \, m \, a^{(\kappa)}_{3}  
{}+ \g^5 \, \frac{m \, \Delta \cdot v'}{\bar P^+} \, a^{(\kappa)}_{4}
{}+ \Pbarslash \, a^{(\kappa)}_{5}
\nn\\[1ex]
&+&
\Pbarslash \, \frac{\Delta \cdot v'}{\bar P^+} \,a^{(\kappa)}_{6}
{}+ \kbarslash \,a^{(\kappa)}_{7}
{}+\kbarslash \, \frac{\Delta \cdot v'}{\bar P^+} \,a^{(\kappa)}_{8}
{}+\Deltaslash \,a^{(\kappa)}_{9}
{}+ \Deltaslash \, \frac{\Delta \cdot v'}{\bar P^+} \,a^{(\kappa)}_{10}
{}+ \vpslash \, \frac{m^2}{P^+} \,a^{(\kappa)}_{11}
\nn\\[1ex]
&+&
\g^5 \, \Pbarslash \,a^{(\kappa)}_{12}
{}+  \g^5 \, \Pbarslash \, \frac{\Delta \cdot v'}{\bar P^+} \,a^{(\kappa)}_{13}
{}+  \g^5 \, \kbarslash \,a^{(\kappa)}_{14}
{}+\g^5 \, \kbarslash \, \frac{\Delta \cdot v'}{\bar P^+} \,a^{(\kappa)}_{15}
{}+ \g^5 \, \Deltaslash \,a^{(\kappa)}_{16}
\nn\\[1ex]
&+&
\g^5 \, \Deltaslash \, \frac{\Delta \cdot v'}{\bar P^+} \, a^{(\kappa)}_{17}
{}+ \g^5 \vpslash \, \frac{m^2}{P^+} \,a^{(\kappa)}_{18}
{}+  \frac{\sigma_{\mu\nu}\bar P^\mu\bar k^\nu}{m} \,a^{(\kappa)}_{19}
{}+   \frac{\sigma_{\mu\nu}\bar P^\mu\bar k^\nu}{m} \, \frac{\Delta \cdot v'}{\bar P^+} \,a^{(\kappa)}_{20}
\nn\\[1ex]
&+&
\frac{\sigma_{\mu\nu}\bar P^\mu\Delta^\nu}{m} \,a^{(\kappa)}_{21}
{}+ \frac{\sigma_{\mu\nu}\bar P^\mu  \Delta^\nu}{m} \,\frac{\Delta \cdot v'}{\bar P^+} \,a^{(\kappa)}_{22}
{}+    \frac{\sigma_{\mu\nu}\Delta^\mu \bar k^\nu}{m} \,a^{(\kappa)}_{23}
{}+   \frac{\sigma_{\mu\nu}  \Delta^\mu\bar k^\nu}{m} \, \frac{\Delta \cdot v'}{\bar P^+} \,a^{(\kappa)}_{24}
\nn\\[1ex]
&+&
\sigma_{\mu\nu}\bar P^\mu v'^{\nu} \frac{m}{\bar P^+} \, a^{(\kappa)}_{25}
{}+\sigma_{\mu\nu}\bar k^\mu v'^{\nu} \frac{m}{\bar P^+} \, a^{(\kappa)}_{26}
{}+ \sigma_{\mu\nu}\Delta^\mu v'^{\nu} \frac{m}{\bar P^+} \, a^{(\kappa)}_{27}
{}+ \frac{\g^5\,\sigma_{\mu\nu}\bar P^\mu \bar k^\nu}{m} \, a^{(\kappa)}_{28}
\nn\\[1ex]
&+&
\frac{\g^5\,\sigma_{\mu\nu}\bar P^\mu \bar k^\nu}{m} \, \frac{\Delta \cdot v'}{\bar P^+} \,a^{(\kappa)}_{29}
{}+\frac{\g^5\,\sigma_{\mu\nu}\bar P^\mu\Delta^\nu}{m} \,a^{(\kappa)}_{30}
{}+\frac{\g^5\,\sigma_{\mu\nu}\bar P^\mu \Delta^\nu}{m} \, \, \frac{\Delta \cdot v'}{\bar P^+} \,a^{(\kappa)}_{31}
\nn\\[1ex]
&+&
\frac{\g^5\,\sigma_{\mu\nu}\Delta^\mu \bar k^\nu}{m} \,a^{(\kappa)}_{32}
{}+ \frac{\g^5\,\sigma_{\mu\nu}\Delta^\mu \bar k^\nu}{m} \, \, \frac{\Delta \cdot v'}{\bar P^+} \, a^{(\kappa)}_{33}
{}+ \g^5\,\sigma_{\mu\nu}\bar P^\mu v'^{\nu} \frac{m}{\bar P^+} \,a^{(\kappa)}_{34}
\nn\\[1ex]
&+&
\g^5\,\sigma_{\mu\nu}\bar k^\mu v'^{\nu} \frac{m}{\bar P^+} \, a^{(\kappa)}_{35}
{}+ \g^5\,\sigma_{\mu\nu}\Delta^\mu v'^{\nu} \frac{m}{\bar P^+} \, a^{(\kappa)}_{36}
{}+    \frac{\g_\mu\,\eps^{\mu \bar P\bar k\Delta}}{m^2} \, a^{(\kappa)}_{37}
{}+ \frac{\g_\mu\,\eps^{\mu \bar P\bar k\Delta}}{m^2} \, \frac{\Delta \cdot v'}{\bar P^+} \,a^{(\kappa)}_{38}
\nn\\[1ex]
&+&
\frac{\g_\mu\,\eps^{\mu v' \bar k \bar P}}{\bar P^+} \,a^{(\kappa)}_{39}
{}+ \frac{\g_\mu\,\eps^{\mu\bar P\Delta v'}}{\bar P^+} \,a^{(\kappa)}_{40}
{}+ \frac{\g_\mu \,\eps^{\mu\bar k\Delta v'}}{\bar P^+} \, a^{(\kappa)}_{41}
{}+    \frac{\g^5 \g_\mu\,\eps^{\mu\bar P\bar k\Delta}}{m^2} \,a^{(\kappa)}_{42}
\nn\\[1ex]
&+&
\frac{\g^5 \g_\mu\,\eps^{\mu\bar P\bar k\Delta}}{m^2} \,\frac{\Delta \cdot v'}{\bar P^+} \, a^{(\kappa)}_{43}
{}+\frac{\g^5 \g_\mu\,\eps^{\mu v' \bar k \bar P }}{\bar P^+} \,a^{(\kappa)}_{44}
{}+ \frac{\g^5 \g_\mu\,\eps^{\mu\bar P\Delta v'}}{\bar P^+} \,a^{(\kappa)}_{45}
{}+ \frac{\g^5 \g_\mu \,\eps^{\mu\bar k\Delta v'}}{\bar P^+} \, a^{(\kappa)}_{46}
\nn\\[1ex]
&+&
\frac{\eps^{\bar P\bar k\Delta v'}}{m^2\,\bar P^+} \Bigg\{ m\, a^{(\kappa)}_{47}
{}+ \g^5\,m\, a^{(\kappa)}_{48}
{}+ \Pbarslash \, a^{(\kappa)}_{49}
{}+ \kbarslash \, a^{(\kappa)}_{50}
{}+\Deltaslash \,a^{(\kappa)}_{51}
{}+ \g^5 \, \Pbarslash \,a^{(\kappa)}_{52}
\nn\\[1ex]
&+&
\g^5 \, \kbarslash \,a^{(\kappa)}_{53}
{}+ \g^5 \, \Deltaslash \, a^{(\kappa)}_{54}
{}+    \frac{\sigma_{\mu\nu}\bar P^\mu \bar k^\nu}{m} \,a^{(\kappa)}_{55}
{}+ \frac{\sigma_{\mu\nu}\bar P^\mu\Delta^\nu}{m} \,a^{(\kappa)}_{56}
{}+ \frac{\sigma_{\mu\nu}\Delta^\mu \bar k^\nu}{m} \,a^{(\kappa)}_{57}
 \Bigg \}
\Bigg] 
\eqpt
\end{eqnarray*} 
\begin{equation}
\label{ansatzfinalmenteold}
\end{equation}
For the helicity non-flipped case we implement the constraints on the non-forward quark-quark correlator 
imposed by the hermiticity properties of the quark fields and their 
well-known behavior under parity and time reversal operations as stated in Eq.~(\ref{constraints}). 
In particular parity invariance imposes the following relations  
\begin{eqnarray*}
a_m^{\kappa} =  a_m^{\kappa} \qquad
m & = &  1,2,5,6,7,8,9,10,11,19,20,21,22,23,\nn\\
&& 24,25,26,27,42,43,44,45,46,48,52,53,54 \nn\\
a_m^{\kappa} = - a_m^{\kappa} \qquad
m & = & 3,4,12,13,14,15,16,17,18,28,29,30,31,32,33,34\nn \\
&& 35,36,37,38,39,40,41,47,49,50,51,55,56,57  
\end{eqnarray*}
\begin{equation}
\label{relazionid1d4_parity}
\end{equation}
Assuming hermiticity  
\begin{eqnarray*}
a_m^{\kappa} = ({a_m^{\kappa}})^* \qquad
m &=& 1,4,5,6,7,10,11,12,14,17,18,19,22,24,25,26, \nn\\
&& 29,30,32,36,37,39,42,45,46,48,51,54,56,57 \nn\\
a_m^{\kappa} =  - ({a_m^{\kappa}})^* \qquad
m & = & 2,3,6,8,9,13,15,16,20,21,22,23,27,28,31,\nn \\
&& 33,34,35,38,40,41,43,44,47,49,50,52,53,55
\end{eqnarray*}
\begin{equation}
\label{relazionid1d4_hermiticity}
\end{equation}
Imposing the time reversal constraint reduces the number of independent amplitudes in the ansatz since 
\begin{eqnarray*}
a_m^{\kappa} = ({a_m^{\kappa}})^* \qquad
m &=& 1,2,3,4,5,6,7,8,9,10,11,12 \nn\\
&&  13,14,15,16,17,18,55,56,57 \nn\\
a_m^{\kappa} =  - ({a_m^{\kappa}})^* \qquad
m &=& 19,20,21,22,23,24,25,26,27,28,29,30, \nn\\
&& 31,32,33,34,35,36,37,38,39,40,41,42 \nn \\
&& 43,44,45,46,47,48,49,50,51,52,53,54
\end{eqnarray*}
\begin{equation}
\label{relazionid1d4_timereversal}
\end{equation}
From Eqs.~(\ref{relazionid1d4_parity}), (\ref{relazionid1d4_hermiticity}) and (\ref{relazionid1d4_timereversal})
the diagonal amplitudes $a_m^{\kappa}$ are either real or pure imaginary. For the case in which the hadron helicity is conserved, (~$\kappa = 1,4$~), 
the ansatz for the quark-quark correlation function can be defined as real by multiplying 
the pure imaginary amplitudes $a_m^{\kappa}$ by the imaginary unity $i$
\begin{eqnarray*}
\Phi^{(\kappa)}(~\bar k,\bar P, \Delta~) &=& \frac{(\delta_{\kappa 1}+\delta_{\kappa 4})}{\sqrt{1-\xi^2}}
 \Bigg [m \,a^{(\kappa)}_{1} 
{}+ \g^5 \, \frac{m \, \Delta \cdot v'}{\bar P^+} \, a^{(\kappa)}_{4} 
{}+ \Pbarslash \, a^{(\kappa)}_{5}
{}+ \kbarslash\,a^{(\kappa)}_{7}
{}+ \Deltaslash \, \frac{\Delta \cdot v'}{\bar P^+} \,a^{(\kappa)}_{10}
\nn\\[1ex]
&+& 
\vpslash \, \frac{m^2}{P^+} \, a^{(\kappa)}_{11} 
{}+ \g^5 \, \Pbarslash \, a^{(\kappa)}_{12} 
{}+\g^5 \, \kbarslash \, a^{(\kappa)}_{14}
{} + \g^5 \, \Deltaslash \, \frac{\Delta \cdot v'}{\bar P^+} \,a^{(\kappa)}_{17} 
{}+ \g^5 \vpslash \, \frac{m^2}{P^+} \,a^{(\kappa)}_{18}
\nn\\[1ex]
&+&
i \,  \frac{\sigma_{\mu\nu}\bar P^\mu\bar k^\nu}{m} \, \frac{\Delta \cdot v'}{\bar P^+} \,a^{(\kappa)}_{20}
{}+ i \,\frac{\sigma_{\mu\nu}\bar P^\mu\Delta^\nu}{m} \,a^{(\kappa)}_{21}  
{}+i \,\frac{\sigma_{\mu\nu}\Delta^\mu \bar k^\nu}{m} \,a^{(\kappa)}_{23}
{} + i \,  \sigma_{\mu\nu}\Delta^\mu v'^{\nu} \frac{m}{\bar P^+} \, a^{(\kappa)}_{27}
\nn\\[1ex]
&+&
i \, \frac{\g^5\,\sigma_{\mu\nu}\bar P^\mu \bar k^\nu}{m} \,a^{(\kappa)}_{28}
{}+ i \, \frac{\g^5\,\sigma_{\mu\nu}\bar P^\mu \Delta^\nu}{m} \,\, \frac{\Delta \cdot v'}{\bar P^+} \,a^{(\kappa)}_{31}
{}+i \, \frac{\g^5\,\sigma_{\mu\nu}\Delta^\mu \bar k^\nu}{m} \,\, \frac{\Delta \cdot v'}{\bar P^+} \, a^{(\kappa)}_{33}
\nn\\[1ex]
&+&
i \, \g^5\,\sigma_{\mu\nu}\bar P^\mu v'^{\nu} \frac{m}{\bar P^+} \, a^{(\kappa)}_{34}
{}+ i \, \g^5\,\sigma_{\mu\nu}\bar k^\mu v'^{\nu} \frac{m}{\bar P^+} \,a^{(\kappa)}_{35}
{}+ i \, \frac{\g_\mu\,\eps^{\mu \bar P\bar k\Delta}}{m^2} \, a^{(\kappa)}_{37}
{}+i \, \frac{\g_\mu\,\eps^{\mu\bar P\Delta v'}}{\bar P^+} \, a^{(\kappa)}_{40}
\nn\\[1ex]
&+&
i \,\frac{\g_\mu \,\eps^{\mu\bar k\Delta v'}}{\bar P^+} \, a^{(\kappa)}_{41}
{}+ \frac{\g^5 \g_\mu\,\eps^{\mu\bar P\bar k\Delta}}{m^2} \, a^{(\kappa)}_{42}
{}+ \frac{\g^5 \g_\mu\,\eps^{\mu\bar P\Delta v'}}{\bar P^+} \, a^{(\kappa)}_{45}
{}+ \frac{\g^5 \g_\mu \,\eps^{\mu\bar k\Delta v'}}{\bar P^+} \, a^{(\kappa)}_{46}
\nn\\[1ex]
&+& 
i \, \frac{\eps^{\bar P\bar k\Delta v'}}{m^2\,\bar P^+}  \, \Bigg \{ m\, a^{(\kappa)}_{47}
{}+ \Pbarslash \,a^{(\kappa)}_{49}
{}+ \kbarslash \, a^{(\kappa)}_{50}
{}+ \g^5 \, \Pbarslash \,a^{(\kappa)}_{52}
{}+ \g^5 \, \kbarslash \,a^{(\kappa)}_{53}
{}+\frac{\sigma_{\mu\nu}\bar P^\mu\Delta^\nu}{i \, m} \, a^{(\kappa)}_{56} 
\nn\\[1ex]
&+&
\frac{\sigma_{\mu\nu}\Delta^\mu \bar k^\nu}{i \, m} \, a^{(\kappa)}_{57}
 \Bigg \}
\Bigg] 
\eqpt
\end{eqnarray*} 
\begin{equation}
\label{ansatzfinalmente}
\end{equation}
If the hadron helicity is flipped from Eqs.~(\ref{eqn:prodotti2}) and 
(\ref{eqn:prodotti3}) one deduces that the ansatz has a factor $-\eta$ 
($\eta^*$), given in Eq.~(\ref{eqnbareta-1}), which reflects the difference in 
phase of the initial and final hadronic spin. From the method introduced in Eq.~(\ref{eqnprodotto_an})
the ansatz for the off-diagonal components of the quark-quark correlation function (~$\kappa=2,3$~) reads
\begin{eqnarray*}
\Phi^{(\kappa)}(~\bar k,\bar P, \Delta~) &=& \frac{({\eta}^*\delta_{\kappa 2}
      -\eta\delta_{\kappa 3})}{\sqrt{t-t_0} \, \sqrt{1-{\xi}^2}}\Bigg[
 m \,a^{(\kappa)}_1 {} +\frac{m \, \Delta \cdot v' }{\bar P^+} \, a^{(\kappa)}_{2}
{}+  m \, \g^5 \, a^{(\kappa)}_{3} 
{}+ \g^5 \, \frac{m \, \Delta \cdot v'}{\bar P^+} \, a^{(\kappa)}_{4}
{}+ \Pbarslash \, a^{(\kappa)}_{5} 
\nn\\[1ex]
&+& 
\Pbarslash \, \frac{\Delta \cdot v'}{\bar P^+} \, a^{(\kappa)}_{6}
{}+\Pbarslash \, {(\frac{\Delta \cdot v'}{\bar P^+})}^2 \, a^{(\kappa)}_{7}
{}+ \kbarslash \, a^{(\kappa)}_{8} 
{}+ \kbarslash \, \frac{\Delta \cdot v'}{\bar P^+} a^{(\kappa)}_{9} 
{}+  \kbarslash \, {({\frac{\Delta \cdot v'}{\bar P^+}})}^2 \, a^{(\kappa)}_{10}
{}+\Deltaslash \,a^{(\kappa)}_{11}   
\nn\\[1ex]
&+&
\Deltaslash \, \frac{\Delta \cdot v'}{\bar P^+} a^{(\kappa)}_{12} 
{}+\Deltaslash {({\frac{\Delta \cdot v'}{\bar P^+}})}^2  a^{(\kappa)}_{13}
{}+\vpslash \, \frac{m^2}{P^+} \, a^{(\kappa)}_{14}
{}+ \vpslash \, \frac{m^2}{P^+} \, \frac{\Delta \cdot v'}{\bar P^+} \, a^{(\kappa)}_{15}
{}+ \g^5 \, \Pbarslash \,a^{(\kappa)}_{16} 
\Bigg. 
\end{eqnarray*}
\begin{eqnarray*}
\Bigg.
&+& \g^5 \, \Pbarslash \, \frac{\Delta \cdot v'}{\bar P^+} \,a^{(\kappa)}_{17}
{}+ \g^5 \, \Pbarslash \, {(\frac{\Delta \cdot v'}{\bar P^+})}^2 \,a^{(\kappa)}_{18}
{}+\g^5 \, \kbarslash \,a^{(\kappa)}_{19}
{}+\g^5 \, \kbarslash \, \frac{\Delta \cdot v'}{\bar P^+} \,a^{(\kappa)}_{20}
{}+ \g^5 \, \kbarslash \, {(\frac{\Delta \cdot v'}{\bar P^+})}^2 \,a^{(\kappa)}_{21}
\nn\\[1ex]
&+&
 \g^5 \, \Deltaslash \, a^{(\kappa)}_{22} 
{}+\g^5 \, \Deltaslash \, \frac{\Delta \cdot v'}{\bar P^+} \, a^{(\kappa)}_{23} 
{}+  \g^5 \, \Deltaslash \, {(\frac{\Delta \cdot v'}{\bar P^+})}^2 \, a^{(\kappa)}_{24} 
{}+ \g^5 \vpslash \, \frac{m^2}{P^+} \, a^{(\kappa)}_{25}
\nn\\[1ex]
&+&
\g^5 \vpslash \,\frac{\Delta \cdot v'}{\bar P^+} \, \frac{m^2}{P^+} \, a^{(\kappa)}_{26}
{}+\frac{\sigma_{\mu\nu}\bar P^\mu\bar k^\nu}{m} \, a^{(\kappa)}_{27}
{}+\frac{\sigma_{\mu\nu}\bar P^\mu\bar k^\nu}{m} \,\frac{\Delta \cdot v'}{\bar P^+} \, a^{(\kappa)}_{28}
{}+\frac{\sigma_{\mu\nu}\bar P^\mu\bar k^\nu}{m} \,{(\frac{\Delta \cdot v'}{\bar P^+})}^2 \, a^{(\kappa)}_{29}
\nn\\[1ex]
&+&
\frac{\sigma_{\mu\nu}\bar P^\mu\Delta^\nu}{m} \, a^{(\kappa)}_{30}
{}+\frac{\sigma_{\mu\nu}\bar P^\mu \Delta^\nu}{m} \,\frac{\Delta \cdot v'}{\bar P^+} \, a^{(\kappa)}_{31}
{}+\frac{\sigma_{\mu\nu}\bar P^\mu \Delta^\nu}{m} \,{(\frac{\Delta \cdot v'}{\bar P^+})}^2 \, a^{(\kappa)}_{32}
{}+\frac{\sigma_{\mu\nu}\bar k^\mu\Delta^\nu}{m} \, a^{(\kappa)}_{33}
\nn\\[1ex]
&+&
\frac{\sigma_{\mu\nu}\bar k^\mu \Delta^\nu}{m} \,\frac{\Delta \cdot v'}{\bar P^+} \, a^{(\kappa)}_{34}
{}+\frac{\sigma_{\mu\nu}\bar k^\mu \Delta^\nu}{m} \,{(\frac{\Delta \cdot v'}{\bar P^+})}^2 \, a^{(\kappa)}_{35}
{}+\sigma_{\mu\nu}\bar P^\mu v'^{\nu} \frac{m}{\bar P^+} \, a^{(\kappa)}_{36}
\nn\\[1ex]
&+&
\sigma_{\mu\nu}\bar P^\mu v'^{\nu} \frac{m}{\bar P^+} \,{(\frac{\Delta \cdot v'}{\bar P^+})}^2 \, a^{(\kappa)}_{37}
{}+\sigma_{\mu\nu}\bar k^\mu v'^{\nu} \frac{m}{\bar P^+} \, a^{(\kappa)}_{38}
{}+\sigma_{\mu\nu}\bar k^\mu v'^{\nu} \frac{m}{\bar P^+} \, {(\frac{\Delta \cdot v'}{\bar P^+})}^2 \, a^{(\kappa)}_{39}
\nn\\[1ex]
&+&
\sigma_{\mu\nu}\Delta^\mu v'^{\nu} \frac{m}{\bar P^+} \, a^{(\kappa)}_{40}
{}+\sigma_{\mu\nu}\Delta^\mu v'^{\nu} \frac{m}{\bar P^+} \, {\frac{\Delta \cdot v'}{\bar P^+}} \, a^{(\kappa)}_{41}
{}+\g^5 \, \frac{\sigma_{\mu\nu}\bar P^\mu\bar k^\nu}{m} \, a^{(\kappa)}_{42}
\nn\\[1ex]
&+&
\g^5 \, \frac{\sigma_{\mu\nu}\bar P^\mu\bar k^\nu}{m} \,\frac{\Delta \cdot v'}{\bar P^+} \, a^{(\kappa)}_{43}
{}+ \g^5 \, \frac{\sigma_{\mu\nu}\bar P^\mu\bar k^\nu}{m} \,{(\frac{\Delta \cdot v'}{\bar P^+})}^2 \, a^{(\kappa)}_{44}
{}+ \g^5 \, \frac{\sigma_{\mu\nu}\bar P^\mu\Delta^\nu}{m} \, a^{(\kappa)}_{45}
\nn\\[1ex]
&+&
\g^5 \, \frac{\sigma_{\mu\nu}\bar P^\mu \Delta^\nu}{m} \,\frac{\Delta \cdot v'}{\bar P^+} \, a^{(\kappa)}_{46}
{}+\g^5 \,\frac{\sigma_{\mu\nu}\bar P^\mu \Delta^\nu}{m} \,{(\frac{\Delta \cdot v'}{\bar P^+})}^2 \, a^{(\kappa)}_{47}
{}+\g^5 \,\frac{\sigma_{\mu\nu}\bar k^\mu\Delta^\nu}{m} \, a^{(\kappa)}_{48}
\nn\\[1ex]
&+&
\g^5 \,\frac{\sigma_{\mu\nu}\bar k^\mu \Delta^\nu}{m} \,\frac{\Delta \cdot v'}{\bar P^+} \, a^{(\kappa)}_{49}
{}+\g^5 \,\sigma_{\mu\nu}\bar P^\mu v'^{\nu} \frac{m}{\bar P^+} \, a^{(\kappa)}_{50}
{}+\g^5 \, \sigma_{\mu\nu}\bar P^\mu v'^{\nu} \frac{m}{\bar P^+} \,{(\frac{\Delta \cdot v'}{\bar P^+})} \, a^{(\kappa)}_{51}
\nn\\[1ex]
&+&
\g^5 \, \sigma_{\mu\nu}\bar k^\mu v'^{\nu} \frac{m}{\bar P^+} \, a^{(\kappa)}_{52}
{}+\g^5 \,\sigma_{\mu\nu}\bar k^\mu v'^{\nu} \frac{m}{\bar P^+} \, {(\frac{\Delta \cdot v'}{\bar P^+})} \, a^{(\kappa)}_{53}
{}+\g^5 \,\sigma_{\mu\nu}\Delta^\mu v'^{\nu} \frac{m}{\bar P^+} \, a^{(\kappa)}_{54}
\nn\\[1ex]
&+&
\g^5 \,\sigma_{\mu\nu}\Delta^\mu v'^{\nu} \frac{m}{\bar P^+} \, 
{\frac{\Delta \cdot v'}{\bar P^+}} \, a^{(\kappa)}_{55}
{}+\frac{\g_\mu\,\eps^{\mu \bar P\bar k\Delta}}{m^2} \, a^{(\kappa)}_{56} 
{}+ \frac{\g_\mu\,\eps^{\mu \bar P\bar k\Delta}}{m^2} \,\frac{\Delta \cdot v'}{\bar P^+} 
\, a^{(\kappa)}_{57} 
{}+\frac{\g_\mu\,\eps^{\mu\bar P\bar k v'}}{\bar P^+} \, a^{(\kappa)}_{58}
\nn\\[1ex]
&+&  
\frac{\g_\mu\,\eps^{\mu\bar P\bar k v'}}{\bar P^+} \, \frac{\Delta \cdot v'}{\bar P^+} 
\, a^{(\kappa)}_{59}
{}+ \frac{\g_\mu\,\eps^{\mu\bar P\Delta v'}}{\bar P^+} \, a^{(\kappa)}_{60}
{}+\frac{\g_\mu\,\eps^{\mu\bar P\Delta v'}}{\bar P^+} \, \frac{\Delta \cdot v'}{\bar P^+} \, a^{(\kappa)}_{61}
{}+\frac{\g_\mu\,\eps^{\mu\bar k\Delta v'}}{\bar P^+} \, a^{(\kappa)}_{62}
\nn\\[1ex]
&+&
\g^5 \, \frac{\g_\mu\,\eps^{\mu \bar P\bar k\Delta}}{m^2} \, a^{(\kappa)}_{63} 
{}+ \g^5 \,\frac{\g_\mu\,\eps^{\mu \bar P\bar k\Delta}}{m^2} \,\frac{\Delta \cdot v'}{\bar P^+} \, a^{(\kappa)}_{64} 
{}+\g^5 \,\frac{\g_\mu\,\eps^{\mu\bar P\bar k v'}}{\bar P^+} \, a^{(\kappa)}_{65}
{}+\g^5 \,\frac{\g_\mu\,\eps^{\mu\bar P\bar k v'}}{\bar P^+} \, \frac{\Delta \cdot v'}{\bar P^+}
\, a^{(\kappa)}_{66}
\nn\\[1ex]
&+&
\g^5 \,\frac{\g_\mu\,\eps^{\mu\bar P\Delta v'}}{\bar P^+} \, a^{(\kappa)}_{67}
{}+\g^5 \,\frac{\g_\mu\,\eps^{\mu\bar P\Delta v'}}{\bar P^+} \, \frac{\Delta \cdot v'}{\bar P^+} \, a^{(\kappa)}_{68}  
{}+\g^5 \,\frac{\g_\mu\,\eps^{\mu\bar k\Delta v'}}{\bar P^+} \, a^{(\kappa)}_{69}
{}+\frac{\eps^{\bar P \bar k \Delta v'}}{m^2 \,\bar P^+} \, [a^{(\kappa)}_{70} 
\nn\\[1ex]
&+&
\frac{\Delta \cdot v'}{\bar P^+} \, a^{(\kappa)}_{71} 
{}+\Pbarslash \, a^{(\kappa)}_{72} 
{}+\Pbarslash \, \frac{\Delta \cdot v'}{\bar P^+} \, a^{(\kappa)}_{73} 
{}+\kbarslash \,a^{(\kappa)}_{74} 
{}+\kbarslash \, a^{(\kappa)}_{75} 
{}+\Deltaslash \, a^{(\kappa)}_{76} 
{}+\Deltaslash \, \frac{\Delta \cdot v'}{\bar P^+} \, a^{(\kappa)}_{77} 
\nn\\[1ex]
&+&
\g^5 \, a^{(\kappa)}_{78} 
{}+\g^5 \, \frac{\Delta \cdot v'}{\bar P^+} \, a^{(\kappa)}_{79} 
{}+\g^5 \,\Pbarslash \, a^{(\kappa)}_{80} 
{}+\g^5 \,\Pbarslash \, \frac{\Delta \cdot v'}{\bar P^+} \, a^{(\kappa)}_{81} 
{}+\g^5 \,\kbarslash \, a^{(\kappa)}_{82} \Bigg .
\end{eqnarray*}
\begin{eqnarray*}
\Bigg .
&+&\g^5 \,\kbarslash \, \frac{\Delta \cdot v'}{\bar P^+} \, a^{(\kappa)}_{83} 
{}+ \g^5 \, \Deltaslash \, a^{(\kappa)}_{84} 
{}+\g^5 \, \Deltaslash \, \frac{\Delta \cdot v'}{\bar P^+} \, a^{(\kappa)}_{85} 
{}+ \frac{\sigma_{\mu\nu}\bar P^\mu\bar k^\nu}{m^2} \,a^{(\kappa)}_{86}
{}+\frac{\sigma_{\mu\nu}\bar P^\mu\bar k^\nu}{m^2} \, \frac{\Delta \cdot v'}{\bar P^+}\,a^{(\kappa)}_{87}
\nn\\[1ex]
&+&
\frac{\sigma_{\mu\nu}\bar P^\mu \Delta^\nu}{m^2} \, a^{(\kappa)}_{88}
{}+\frac{\sigma_{\mu\nu}\bar P^\mu \Delta^\nu}{m^2} \, \frac{\Delta \cdot v'}{\bar P^+}\,a^{(\kappa)}_{89}
{}+\frac{\sigma_{\mu\nu}\bar k^\mu \Delta^\nu}{m^2} \,a^{(\kappa)}_{90}
{}+\frac{\sigma_{\mu\nu}\bar k^\mu \Delta^\nu}{m^2} \, \frac{\Delta \cdot v'}{\bar P^+} \,a^{(\kappa)}_{91}
\Bigg] 
\nn\\[1ex]
&+& m \, \left( \frac{{\eta}^*\delta_{\kappa 2}
      -\eta\delta_{\kappa 3}}{\sqrt{t-t_0}} \, \right)\Bigg[
 m \,a^{(\kappa)}_{92} 
{}+m \, \g^5 \, a^{(\kappa)}_{93} 
{}+ \Pbarslash \, a^{(\kappa)}_{94} 
{}+\kbarslash \, a^{(\kappa)}_{95}
{}+\Deltaslash \,a^{(\kappa)}_{96}
{}+ \g^5 \, \Pbarslash \,a^{(\kappa)}_{97} 
{}+\g^5 \, \kbarslash \,a^{(\kappa)}_{98}
\nn\\[1ex]
&+& 
\g^5 \, \Deltaslash \, a^{(\kappa)}_{99}
{}+\frac{\sigma_{\mu\nu}\bar P^\mu\bar k^\nu}{m} \, a^{(\kappa)}_{100}
{}+\frac{\sigma_{\mu\nu}\bar P^\mu\Delta^\nu}{m} \, a^{(\kappa)}_{101}
{}+\frac{\sigma_{\mu\nu}\bar k^\mu\Delta^\nu}{m} \, a^{(\kappa)}_{102}
{}+ \g^5 \, \frac{\sigma_{\mu\nu}\bar P^\mu\bar k^\nu}{m} \, a^{(\kappa)}_{103}
\nn\\[1ex]
&+&
\g^5 \, \frac{\sigma_{\mu\nu}\bar P^\mu\Delta^\nu}{m} \, a^{(\kappa)}_{104}
{}+\g^5 \,\frac{\sigma_{\mu\nu}\bar k^\mu\Delta^\nu}{m} \, a^{(\kappa)}_{105}
{}+\frac{\g_\mu\,\eps^{\mu \bar P\bar k\Delta}}{m^2} \, a^{(\kappa)}_{106} 
{}+ \g^5 \, \frac{\g_\mu\,\eps^{\mu \bar P\bar k\Delta}}{m^2} \, a^{(\kappa)}_{107}
\Bigg] 
\nn\\[1ex]
&+&
\left({\eta}^*\delta_{\kappa 2}
      -\eta\delta_{\kappa 3}  \right)\, (\frac{\sqrt{t-t_0}}{m})  \, \Bigg[
m \,a^{(\kappa)}_{108} 
{}+  m \, \g^5 \, a^{(\kappa)}_{109} 
{}+ \Pbarslash \, a^{(\kappa)}_{110} 
{}+ \kbarslash \, a^{(\kappa)}_{111}
{}+\Deltaslash \,a^{(\kappa)}_{112}
{}+\g^5 \, \Pbarslash \,a^{(\kappa)}_{113} 
\nn\\[1ex]
&+&
\g^5 \, \kbarslash \,a^{(\kappa)}_{114}
{}+\g^5 \, \Deltaslash \, a^{(\kappa)}_{115} 
{}+\frac{\sigma_{\mu\nu}\bar P^\mu\bar k^\nu}{m} \, a^{(\kappa)}_{116}
{}+\frac{\sigma_{\mu\nu}\bar P^\mu\Delta^\nu}{m} \, a^{(\kappa)}_{117}
{}+\frac{\sigma_{\mu\nu}\bar k^\mu\Delta^\nu}{m} \, a^{(\kappa)}_{118}
\nn\\[1ex]
&+&
\g^5 \, \frac{\sigma_{\mu\nu}\bar P^\mu\bar k^\nu}{m} \, a^{(\kappa)}_{119}
{}+ \g^5 \, \frac{\sigma_{\mu\nu}\bar P^\mu\Delta^\nu}{m} \, a^{(\kappa)}_{120}
{}+\g^5 \,\frac{\sigma_{\mu\nu}\bar k^\mu\Delta^\nu}{m} \, a^{(\kappa)}_{121}
{}+ \frac{\g_\mu\,\eps^{\mu \bar P\bar k\Delta}}{m^2} \, a^{(\kappa)}_{122} 
{}+ \g^5 \, \frac{\g_\mu\,\eps^{\mu \bar P\bar k\Delta}}{m^2} \, a^{(\kappa)}_{123}
\Bigg]
\eqpt
\end{eqnarray*} 
\begin{equation} 
\label{ansatzfinalmenteoff} 
\end{equation}
We implement again hermiticity, parity and time reversal invariance. In particular parity invariance 
imposes the following relations  
\begin{eqnarray*}
d_m^{2} = {\eta}^2 \, d_m^{3} \qquad
m & = &  1,2,5,6,7,8,9,10,11,12,13,14,15,27,28,29,30,31,\nn\\
&& 32,33,34,35,36,37,38,39,40,41,63,64,65,66,67,68,\nn\\
&& 69,78,79,80,81,82,83,84,85,92,94,95,96,100,101,\nn\\
&&102,107,108,110,111,112,116,117,118,123 
\nn\\[2mm]
d_m^{2} = - {\eta}^2 \, d_m^{3} \qquad
m & = & 3,4,16,17,18,19,20,21,22,22,23,24,25,26,42,43,\nn \\
&& 44,45,46,47,48,49,50,51,52,53,54,55,56,57,58,59,\nn\\
&& 60,61,62,70,71,72,73,74,75,76,77,86,87,88,89,90,\nn \\
&&91,93,97,98,99,103,104,105,106,109,113,114,115,119,\nn \\
&&120,121,122, \, .
\end{eqnarray*}
\begin{equation}
\label{relazionid2d3_parity}
\end{equation}
Applying hermiticity   
\begin{eqnarray*}
d_m^{2} = ({d_m^{3}})^* \qquad
m &=& 1,4,5,7,8,10,12,14,16,18,19,21,23,25,27,29,31,   \nn\\
&& 34,36,37,38,39,41,43,45,47,48,51,53,54,57,58,61,  \nn\\
&& 64,65,68,73,75,76,78,79,81,83,84,87,88,90,92,94, \nn \\
&&95,97,98,100,104,105,108,110,111,113,114,116,120,121
\nn\\[2mm]
d_m^{2} =  - ({d_m^{3}})^* \qquad
m &=& 2,3,6,9,11,13,15,17,20,22,24,26,28,30,32,33,\nn\\
&& 35,40,42,44,46,49,50,52,55,56,59,60,62,63,66,67,\nn\\
&& 69,70,71,72,74,77,80,82,85,86,89,91,93,96,99,101,\nn\\
&&102,103,106,107,109,112,115,117,118,119,122,123 \, .
\end{eqnarray*}
\begin{equation}
\label{relazionid2d3_hermiticity}
\end{equation}
Imposing the time reversal constraint reduces the number of independent amplitudes in the ansatz since 
\begin{eqnarray*}
d_m^{\kappa} = {\eta}^2 \, ({d_m^{\kappa}})^* \qquad
m &=& 27,28,29,30,31,32,33,34,35,36,37,38,39,40,41,\nn\\
&& 42,43,44,45,46,47,48,49,50,51,52,53,54,55,56,57,\nn\\
&&58,59,60,61,62,63,64,65,66,67,68,69,70,71,72,73,74, \nn \\
&& 75,76,77,78,79,80,81,82,83,84,85,100,101,102,103,\nn\\
&&104,105,106,107,116,117,118,119,120,121,122,123 \nn\\
d_m^{\kappa} = - {\eta}^2 \, ({d_m^{\kappa}})^* \qquad
m &=& 1,2,3,4,5,6,7,8,9,10,11,12,13,14,15,16,17,18 \nn \\
&&19,20,21,22,23,24,25,26,86,87,88,89,90,91,92,93,94,\nn \\
&&95,96,97,98,99,108,109,110,111,112,113,114,115, \, 
\end{eqnarray*}
\begin{equation}
\label{relazionid2d3_timereversal}
\end{equation}
Because of the relations in Eq.~(\ref{relazionid2d3_parity}), (\ref{relazionid2d3_hermiticity}) and 
(\ref{relazionid2d3_timereversal}) some of the amplitudes are zero. We refrain from rewriting the lengthy 
expression of the off-diagonal ansatz.  

In the forward limit $\Delta=0$ the spinorial products in Eq.(\ref{eqn:prodotti2}) and Eq.(\ref{eqn:prodotti3}) 
cannot be built through the trace method since the product  $\bar u(P,-)\,u(P,+)=\bar u(P,+)\,u(P,-)=0$. This implies 
that the off-diagonal part of the ansazt does not converge. The method here developed 
is therefore applicable only to the cases for which $\Delta$ is different from zero.

\begin{acknowledgments}
We would like to thank Dr.~Markus Diehl, Dr.~Alessandro Bacchetta 
and, in particular, Dr.~Rainer Jakob for their helpful discussions 
and support. 
\end{acknowledgments}


\end{document}